\documentclass{cernyrep}
\newcommand{\re}[1]{\ensuremath{\mathcal{R}e(#1)}}
\newcommand{\im}[1]{\ensuremath{\mathcal{I}m(#1)}}
\newcommand{\Bz}{{B^0}}
\newcommand{\Bzb}{\overline{B}{}^0}

\newcommand{\CP}{CP\ }

\newcommand{\Bbar}{\overline{B}}

\newcommand{\Heff}{\mathcal{H}}
\newcommand{\Meff}{M}
\newcommand{\Geff}{\Gamma}

\newcommand{\fb}{\overline{f}}
\newcommand{\f}{f}
\def\lsim{\mathrel{\rlap{\lower4pt\hbox{\hskip1pt$\sim$}}
    \raise1pt\hbox{$<$}}}         
\def\gsim{\mathrel{\rlap{\lower4pt\hbox{\hskip1pt$\sim$}}
    \raise1pt\hbox{$>$}}}         

\begin{document}

\title{Flavour physics and CP violation}

\author{Y. Nir}
\institute{Weizmann Institute of Science, Rehovot, Israel}
\maketitle

\begin{abstract}
  This is a written version of a series of lectures aimed at graduate
  students in particle theory/string theory/particle experiment
  familiar with the basics of the Standard Model. We explain the many
  reasons for the interest in flavour physics. We describe flavour
  physics and the related CP violation within the Standard Model, and
  explain how the B-factories proved that the Kobayashi-Maskawa
  mechanism dominates the CP violation that is observed in meson
  decays.  We explain the implications of flavour physics for new
  physics.  We emphasize the ``new physics flavour puzzle''. As an
  explicit example, we explain how the recent measurements of
  $D^0-\overline{D}^0$ mixing constrain the supersymmetric flavour
  structure.  We explain how the ATLAS and CMS experiments can solve
  the new physics flavour puzzle and perhaps shed light on the standard
  model flavour puzzle. Finally, we describe various interpretations of
  the neutrino flavour data and their impact on flavour models.
  \end{abstract}

\section{What is flavour?}
\label{sec:intro}

The term `\textbf{flavours}' is used, in the jargon of particle
physics, to describe several copies of the same gauge representation,
namely several fields that are assigned the same quantum
charges. Within the Standard Model, when thinking of its unbroken
$SU(3)_\text{C}\times U(1)_\text{EM}$ gauge group, there are four
different types of particles, each coming in three flavours:
\begin{itemize}
\item Up-type quarks in the $(3)_{+2/3}$ representation: $u,c,t$.
\item Down-type quarks in the $(3)_{-1/3}$ representation: $d,s,b$.
\item Charged leptons in the $(1)_{-1}$ representation: $e,\mu,\tau$.
\item Neutrinos in the $(1)_{0}$ representation: $\nu_1,\nu_2,\nu_3$.
\end{itemize}

The term `\textbf{flavour physics}' refers to interactions that
distinguish between flavours. By definition, gauge interactions,
namely interactions that are related to unbroken symmetries and
mediated therefore by massless gauge bosons, do not distinguish among
the flavours and do not constitute part of flavour physics. Within the
Standard Model, flavour physics refers to the weak and Yukawa
interactions.

The term `\textbf{flavour parameters}' refers to parameters that carry
flavour indices. Within the Standard Model, these are the nine masses of
the charged fermions and the four `mixing parameters' (three angles
and one phase) that describe the interactions of the
charged weak-force carriers ($W^\pm$) with quark--antiquark pairs. If
one augments the Standard Model with Majorana mass terms for the
neutrinos, one should add to the list three neutrino masses and six
mixing parameters (three angles and three phases) for the $W^\pm$
interactions for lepton--antilepton pairs.

The term `\textbf{flavour universal}' refers to interactions with couplings
(or to flavour parameters) that are proportional to the unit matrix in
flavour space. Thus, the strong and electromagnetic interactions are
flavour universal\footnote
   {In the interaction basis, the weak interactions are also
    flavour universal, and one can identify the source of all flavour
    physics in the Yukawa interactions among the gauge-interaction
    eigenstates.}.
An alternative term for `flavour universal' is `\textbf{flavour blind}'.

The term `\textbf{flavour diagonal}' refers to interactions with couplings (or
to flavour parameters) that are diagonal, but not necessarily
universal, in the flavour space. Within the Standard Model, the Yukawa
interactions of the Higgs particle are flavour diagonal in the mass
basis.

The term `\textbf{flavour changing}' refers to processes where the
initial and final flavour-numbers (that is, the number of particles of a
certain flavour minus the number of antiparticles of the same flavour)
are different. In `flavour-changing charged current' processes, both
up-type and down-type flavours, and/or both charged lepton and neutrino
flavours are involved. Examples are (i) muon decay via $\mu\to
e\bar\nu_i\nu_j$, and (ii) $K^-\to\mu^-\bar\nu_j$ (which corresponds,
at the quark level, to $s\bar u\to\mu^-\bar\nu_j$). Within the
Standard Model, these processes are mediated by the $W$ bosons and
occur at tree level. In `\textbf{flavour-changing neutral current}' (FCNC)
processes, either up-type or down-type flavours but not both, and/or
either charged lepton or neutrino flavours but not both, are involved.
Examples are (i) muon decay via $\mu\to e\gamma$ and (ii)
$K_L\to\mu^+\mu^-$ (which corresponds, at the quark level, to $s\bar
d\to\mu^+\mu^-$). Within the Standard Model, these processes do not
occur at tree level, and are often highly suppressed.

Another useful term is `\textbf{flavour violation}'. We shall explain
it later in these lectures.

\section{Why is flavour physics interesting?}
\label{sec:mot}

\begin{itemize}
\item Flavour physics can discover new physics or probe it before it
      is directly observed in experiments. Here are some examples from
      the past:
  \begin{itemize}
  \item The smallness of $\frac{\Gamma(K_L\to\mu^+\mu^-)}
        {\Gamma(K^+\to\mu^+\nu)}$ led to the prediction of a fourth
        (the charm) quark.
  \item The size of $\Delta m_K$ led to a successful prediction of the
        charm mass.
  \item The size of $\Delta m_B$ led to a successful prediction of the
        top mass.
  \item The measurement of $\varepsilon_K$ led to the prediction of
        the third generation.
  \end{itemize}
\item CP violation is closely related to flavour physics. Within the
      Standard Model, there is a single CP-violating parameter, the
      Kobayashi--Maskawa phase $\delta_\text{KM}$
      \cite{Kobayashi:1973fv}.  Baryogenesis tells us, however, that
      there must exist new sources of CP violation. Measurements of CP
      violation in flavour-changing processes might provide evidence
      for such sources.
\item The fine-tuning problem of the Higgs mass, and the puzzle of
      dark matter imply that there exists new physics at, or below,
      the \UTeVZ{} scale. If such new physics had a generic flavour
      structure, it would contribute to flavour-changing neutral
      current (FCNC) processes orders of magnitude above the observed
      rates. The question of why this does not happen constitutes the
      \emph{new physics flavour puzzle}.
\item Most of the charged fermion flavour parameters are small and
  hierarchical. The Standard Model does not provide any explanation of
  these features. This is the \emph{Standard Model flavour puzzle}.  The
  puzzle became even deeper after neutrino masses and mixings were
  measured because, so far, neither smallness nor hierarchy in these
  parameters have been established.
\end{itemize}

\section{Flavour in the Standard Model}
\label{smfor}

A model of elementary particles and their interactions is defined by
the following ingredients: (i) The symmetries of the Lagrangian and
the pattern of spontaneous symmetry breaking; (ii) The representations
of fermions and scalars. The Standard Model (SM) is defined as
follows:\\ (i) The gauge symmetry is
\begin{equation}\label{smsym}
G_\text{SM}=SU(3)_\text{C}\times SU(2)_\text{L}\times U(1)_\text{Y}.
\end{equation}
It is spontaneously broken by the VEV of a single Higgs scalar,
$\phi(1,2)_{1/2}$ ($\langle\phi^0\rangle=v/\sqrt{2}$):
\begin{equation}\label{smssb}
G_\text{SM} \to SU(3)_\text{C}\times U(1)_\text{EM}.
\end{equation}
(ii) There are three fermion generations, each consisting of five
representations of $G_\text{SM}$:
\begin{equation}\label{ferrep}
Q_{Li}(3,2)_{+1/6},\ \ U_{Ri}(3,1)_{+2/3},\ \
D_{Ri}(3,1)_{-1/3},\ \ L_{Li}(1,2)_{-1/2},\ \ E_{Ri}(1,1)_{-1}.
\end{equation}

\subsection{The interactions basis}
The Standard Model Lagrangian, $\mathcal{L}_\text{SM}$, is the most
general renormalizable Lagrangian that is consistent with the gauge
symmetry (\ref{smsym}), the particle content (\ref{ferrep}) and the
pattern of spontaneous symmetry breaking (\ref{smssb}). It can be
divided into three parts:
\begin{equation}\label{LagSM}
\mathcal{L}_\text{SM}=\mathcal{L}_\text{kinetic}+\mathcal{L}_\text{Higgs}
+\mathcal{L}_\text{Yukawa}.
\end{equation}

For the kinetic terms, to maintain gauge invariance, one has to
replace the derivative with a covariant derivative:
\begin{equation}\label{SMDmu}
D^\mu=\partial^\mu+ig_s G^\mu_a L_a+ig W^\mu_b T_b+ig^\prime B^\mu Y.
\end{equation}
Here $G^\mu_a$ are the eight gluon fields, $W^\mu_b$ the three weak
interaction bosons, and $B^\mu$ the single hypercharge boson.  The
$L_a$'s are $SU(3)_\text{C}$ generators (the $3\times3$ Gell-Mann
matrices $\frac{1}{2}\lambda_a$ for triplets, $0$ for singlets), the
$T_b$'s are $SU(2)_\text{L}$ generators (the $2\times2$ Pauli matrices
$\frac{1}{2}\tau_b$ for doublets, $0$ for singlets), and the $Y$'s are
the $U(1)_\text{Y}$ charges. For example, for the quark doublets
$Q_L$, we have
\begin{equation}\label{DmuQL}
\mathcal{L}_\text{kinetic}(Q_L)= i{\overline{Q_{Li}}}\gamma_\mu
\left(\partial^\mu+\frac{i}{2}g_s G^\mu_a\lambda_a
+\frac{i}{2}g W^\mu_b\tau_b+\frac{i}{6}g^\prime
B^\mu\right)\delta_{ij}Q_{Lj},
\end{equation}
while for the lepton doublets $L_L^I$, we have
\begin{equation}\label{DmuLL}
\mathcal{L}_\text{kinetic}(L_L)= i{\overline{L_{Li}}}\gamma_\mu
\left(\partial^\mu+\frac{i}{2}g W^\mu_b\tau_b-\frac i2 g^\prime
  B^\mu\right)\delta_{ij}L_{Lj}.
\end{equation}
The unit matrix in flavour space, $\delta_{ij}$, signifies that
these parts of the interaction Lagrangian are flavour universal. In
addition, they conserve CP.

The Higgs potential, which describes the scalar self-interactions, is
given by
\begin{equation}\label{HiPo}
\mathcal{L}_\text{Higgs}=\mu^2\phi^\dagger\phi-\lambda(\phi^\dagger\phi)^2.
\end{equation}
For the Standard Model scalar sector, where there is a single doublet,
this part of the Lagrangian is also CP conserving.

The quark Yukawa interactions are given by
\begin{equation}\label{Hqint}
-\mathcal{L}_\text{Y}^{q}=Y^d_{ij}{\overline {Q_{Li}}}\phi D_{Rj}
+Y^u_{ij}{\overline {Q_{Li}}}\tilde\phi U_{Rj}+\text{h.c.},
\end{equation}
(where $\tilde\phi=i\tau_2\phi^\dagger$) while the lepton Yukawa
interactions are given by
\begin{equation}\label{Hlint}
-\mathcal{L}_\text{Y}^{\ell}=Y^e_{ij}{\overline {L_{Li}}}\phi E_{Rj}
+\text{h.c.}
\end{equation}
This part of the Lagrangian is, in general, flavour dependent (that is,
$Y^f\not\propto\mathbf{1}$) and CP violating.

\subsection{Global symmetries}
\label{sec:spurions}
In the absence of the Yukawa matrices $Y^d$, $Y^u$ and $Y^e$, the SM
has a large $U(3)^5$ global symmetry:
\begin{equation}\label{gglobal}
G_\text{global}(Y^{u,d,e}=0)=SU(3)_q^3\times SU(3)_\ell^2\times U(1)^5,
\end{equation}
where
\begin{eqnarray}\label{susuu}
SU(3)_q^3&=&SU(3)_Q\times SU(3)_U\times SU(3)_D,\nonumber\\
SU(3)_\ell^2&=&SU(3)_L\times SU(3)_E,\nonumber\\
U(1)^5&=&U(1)_B\times U(1)_L\times U(1)_Y\times U(1)_\text{PQ}\times
U(1)_E.
\end{eqnarray}
Out of the five $U(1)$ charges, three can be identified with baryon
number ($B$), lepton number ($L$), and hypercharge ($Y$), which are
respected by the Yukawa interactions. The two remaining $U(1)$ groups
can be identified with the PQ symmetry whereby the Higgs and $D_R,E_R$
fields have opposite charges, and with a global rotation of $E_R$
only.

The point that is important for our purposes is that
$\mathcal{L}_\text{kinetic}+\mathcal{L}_\text{Higgs}$ respect the
non-Abelian flavour symmetry $S(3)_q^3\times SU(3)_\ell^2$, under which
\begin{equation}\label{symkh}
Q_L\to V_QQ_L,\ \ \ U_R\to V_U U_R,\ \ \ D_R\to V_D D_R,\ \ L_L\to V_L
L_L,\ \ \ E_R\to V_E E_R,
\end{equation}
where the $V_i$ are unitary matrices.
The Yukawa interactions (\ref{Hqint}) and (\ref{Hlint}) break the
global symmetry,
\begin{equation}\label{globre}
G_\text{global}(Y^{u,d,e}\neq0)= U(1)_B\times U(1)_e\times
U(1)_\mu\times U(1)_\tau.
\end{equation}
(Of course, the gauged $U(1)_Y$ also remains a good symmetry.)  Thus,
the transformations of \Eref{symkh} are not a symmetry of
$\mathcal{L}_\text{SM}$. Instead, they correspond to a change of the
interaction basis. These observations also offer an alternative way of
defining flavour physics: it refers to interactions that break the
$SU(3)^5$ symmetry (\ref{symkh}). Thus, the term `\textbf{flavour
violation}' is often used to describe processes or parameters that
break the symmetry.

One can think of the quark Yukawa couplings as spurions that break the
global $SU(3)_q^3$ symmetry (but are neutral under $U(1)_B$),
\begin{equation}\label{Gglobq}
Y^u\sim(3,\bar3,1)_{SU(3)_q^3},\ \ \
Y^d\sim(3,1,\bar3)_{SU(3)_q^3},
\end{equation}
and of the lepton Yukawa couplings as spurions that break the global
$SU(3)_\ell^2$ symmetry (but are neutral under $U(1)_e\times
U(1)_\mu\times U(1)_\tau$),
\begin{equation}\label{Gglobl}
Y^e\sim(3,\bar3)_{SU(3)_\ell^2}.
\end{equation}
The spurion formalism is convenient for several purposes: parameter
counting (see below), identification of flavour suppression factors
(see \Sref{sec:nppuzzle}), and the idea of minimal flavour
violation (see \Sref{sec:lhc}).

\subsection{Counting parameters}

How many independent parameters are there in $\mathcal{L}_\text{Y}^q$?
The two Yukawa matrices, $Y^u$ and $Y^d$, are $3\times3$ and complex.
Consequently, there are 18 real and 18 imaginary parameters in these
matrices. Not all of them are, however, physical. The pattern of
$G_\text{global}$ breaking means that there is freedom to remove 9
real and 17 imaginary parameters (the number of parameters in three
$3\times3$ unitary matrices minus the phase related to $U(1)_B$). For
example, we can use the unitary transformations $Q_L\to V_QQ_L$,
$U_R\to V_U U_R$, and $D_R\to V_D D_R$ to lead to the following
interaction basis:
\begin{equation}\label{speint}
Y^d=\lambda_d,\ \ \ Y^u=V^\dagger\lambda_u,
\end{equation}
where $\lambda_{d,u}$ are diagonal,
\begin{equation}\label{deflamd}
\lambda_d=\text{diag}(y_d,y_s,y_b),\ \ \
\lambda_u=\text{diag}(y_u,y_c,y_t),
\end{equation}
while $V$ is a unitary matrix that depends on three real angles and
one complex phase. We conclude that there are 10 quark flavour
parameters: 9 real ones and a single phase. In the mass basis, we
shall identify the nine real parameters as six quark masses and three
mixing angles, while the single phase is $\delta_\text{KM}$.

How many independent parameters are there in $\mathcal{L}_\text{Y}^\ell$?
The Yukawa matrix $Y^e$ is $3\times3$ and complex. Consequently, there
are 9 real and 9 imaginary parameters in this matrix. There is,
however, freedom to remove 6 real and 9 imaginary parameters (the
number of parameters in two $3\times3$ unitary matrices minus the
phases related to $U(1)^3$). For example, we can use the unitary
transformations $L_L\to V_LL_L$ and $E_R\to V_E E_R$ to lead to the
following interaction basis:
\begin{equation}\label{speintl}
Y^e=\lambda_e=\text{diag}(y_e,y_\mu,y_\tau).
\end{equation}
We conclude that there are three real lepton flavour parameters. In
the mass basis, we shall identify these parameters as the three
charged lepton masses. We must, however, modify the model when we take
into account the evidence for neutrino masses.

\subsection{The mass basis}
Upon the replacement $\re{\phi^0}\to\frac{v+H^0}{\sqrt2}$, the Yukawa
interactions (\ref{Hqint}) give rise to the mass matrices
\begin{equation}\label{YtoMq}
M_q=\frac{v}{\sqrt2}Y^q.
\end{equation}
The mass basis corresponds, by definition, to diagonal mass
matrices. We can  always find unitary matrices $V_{qL}$ and $V_{qR}$
such that
\begin{equation}\label{diagMq}
V_{qL}M_q V_{qR}^\dagger=M_q^\text{diag}\equiv\frac{v}{\sqrt2}\lambda_q.
\end{equation}
The four matrices $V_{dL}$, $V_{dR}$, $V_{uL}$, and $V_{uR}$ are then
the ones required to transform to the mass basis. For example, if we
start from the special basis (\ref{speint}), we have
$V_{dL}=V_{dR}=V_{uR}=\mathbf{1}$ and $V_{uL}=V$. The combination
$V_{uL}V_{dL}^\dagger$ is independent of the interaction basis from
which we start this procedure.

We denote the left-handed quark mass eigenstates as $U_L$ and $D_L$.
The charged-current interactions for quarks [that is the interactions of the
charged $SU(2)_\text{L}$ gauge bosons $W^\pm_\mu=\frac{1}{\sqrt{2}}
(W^1_\mu\mp iW_\mu^2)$], which in the interaction basis are described
by (\ref{DmuQL}), have a complicated form in the mass basis:
\begin{equation}\label{Wmasq}
-\mathcal{L}_{W^\pm}^q=\frac{g}{\sqrt{2}}{\overline {U_{Li}}}\gamma^\mu
V_{ij}D_{Lj} W_\mu^++\text{h.c.}\ ,
\end{equation}
where $V$ is the $3\times3$ unitary matrix ($VV^\dagger=V^\dagger
V=\mathbf{1}$) that appeared in \Eref{speint}. For a general
interaction basis,
\begin{equation}\label{VCKM}
V=V_{uL}V_{dL}^\dagger.
\end{equation}
$V$ is the Cabibbo--Kobayashi--Maskawa (CKM) \emph{mixing matrix} for
quarks \cite{Cabibbo:1963yz,Kobayashi:1973fv}. As a result of the fact
that $V$ is not diagonal, the $W^\pm$ gauge bosons couple to quark
mass eigenstates of different generations. Within the Standard
Model, this is the only source of \emph{flavour-changing} quark
interactions.

\textbf{Exercise 1:} \emph{Prove that, in the absence of neutrino masses, there
is no mixing in the lepton sector.}

\textbf{Exercise 2:} \emph{Prove that there is no mixing in the $Z$
couplings. (In the jargon of physics, there are no flavour-changing
neutral currents at tree level.)}

The detailed structure of the CKM matrix, its parametrization, and the
constraints on its elements are described in Appendix \ref{app:ckm}.

\section{Testing CKM}
\label{sec:bmix}

Measurements of rates, mixing, and CP asymmetries in $B$ decays in the
two B factories, BaBar and Belle, and in the two Tevatron detectors,
CDF and D0, signified a new era in our understanding of CP
violation. The progress is both qualitative and quantitative. Various
basic questions concerning CP and flavour violation have, for the
first time, received answers based on experimental information. These
questions include, for example,
\begin{itemize}
\item Is the Kobayashi--Maskawa mechanism at work (namely, is
  $\delta_\text{KM}\neq0$)?
\item Does the KM phase dominate the observed CP violation?
\end{itemize}
As a first step, one may assume the SM and test the overall
consistency of the various measurements. However, the richness of data
from the B factories allows us to go a step further and answer these
questions model independently, namely allowing new physics to
contribute to the relevant processes. We here explain the way in which
this analysis proceeds.

\subsection{$S_{\psi K_S}$}
The CP asymmetry in $B\to\psi K_S$ decays plays a major role in
testing the KM mechanism. Before we explain the test itself, we should
understand why the theoretical interpretation of the asymmetry is
exceptionally clean, and what are the theoretical parameters on which
it depends, within and beyond the Standard Model.

The CP asymmetry in neutral meson decays into final CP eigenstates
$f_{\CP}$ is defined as follows:
\begin{equation}\label{asyfcpt}
\mathcal{A}_{f_{\CP}}(t)\equiv\frac{d\Gamma/dt[\Bzb_\text{phys}(t)\to f_{\CP}]-
d\Gamma/dt[\Bz_\text{phys}(t)\to f_{\CP}]}
{d\Gamma/dt[\Bzb_\text{phys}(t)\to f_{\CP}]+d\Gamma/dt[\Bz_\text{phys}(t)\to
  f_{\CP}]}\; .
\end{equation}
A detailed evaluation of this asymmetry is given in Appendix
\ref{sec:formalism}. It leads to the following form:
\begin{eqnarray}\label{asyfcpbt}
\mathcal{A}_{f_{\CP}}(t)&=&S_{f_{\CP}}\sin(\Delta
mt)-C_{f_{\CP}}\cos(\Delta mt),\nonumber\\
S_{f_{\CP}}&\equiv&\frac{2\,\im{\lambda_{f_{\CP}}}}{1+|\lambda_{f_{\CP}}|^2},\
\ \  C_{f_{\CP}}\equiv\frac{1-|\lambda_{f_{\CP}}|^2}{1+|\lambda_{f_{\CP}}|^2}
\; ,
\end{eqnarray}
where
\begin{equation}\label{lamhad}
\lambda_{f_{\CP}}=e^{-i\phi_B}(\overline{A}_{f_{\CP}}/A_{f_{\CP}}) \; .
\end{equation}
Here $\phi_B$ refers to the phase of $M_{12}$ [see
\Eref{defmgam}].  Within the Standard Model, the corresponding
phase factor is given by
\begin{equation}\label{phimsm}
e^{-i\phi_B}=(V_{tb}^* V_{td}^{})/(V_{tb}^{}V_{td}^*) \;.
\end{equation}
The decay amplitudes $A_f$ and $\overline{A}_f$ are defined in
\Eref{decamp}.

\begin{figure}[htb]
\centering
\includegraphics[width=.45\linewidth]{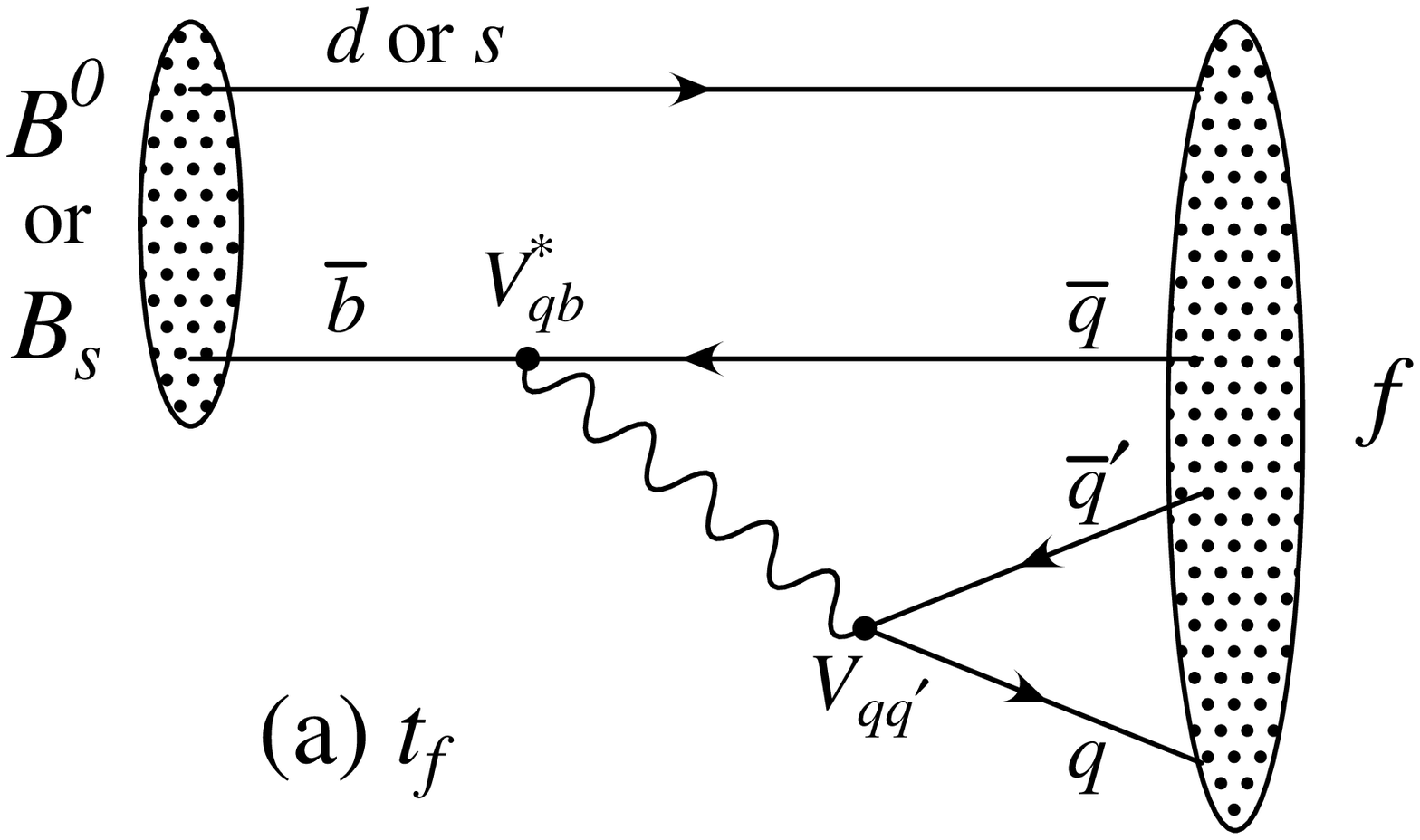}
\quad
\includegraphics[width=.45\linewidth]{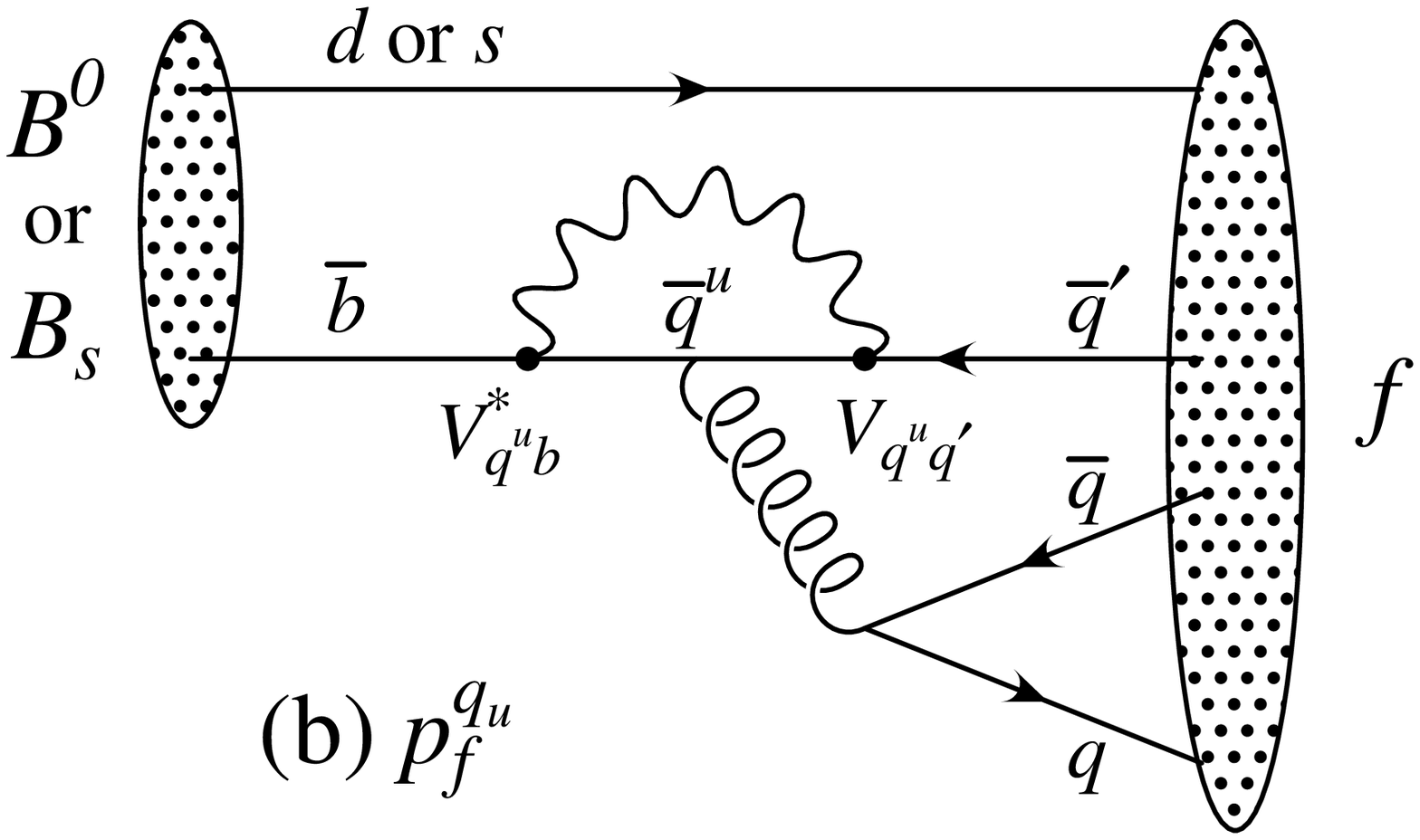}
\caption[]{Feynman diagrams for (a) tree and (b) penguin amplitudes
           contributing to $B^0\to f$ or $B_{s}\to f$ via a $\bar
           b\to\bar q q\bar q^\prime$ quark-level process}
\label{fig:diags}
\end{figure}

The $B^0\to J/\psi K^0$ decay~\cite{Carter:1980hr,Bigi:1981qs}
proceeds via the quark transition $\bar b\to\bar c c\bar s$. There are
contributions from both tree ($t$) and penguin ($p^{q_u}$, where
$q_u=u,c,t$ is the quark in the loop) diagrams (see \Fref{fig:diags})
which carry different weak phases:
\begin{equation}\label{ckmdec}
A_f = \left(V^\ast_{cb} V^{}_{cs}\right) t_f +
\sum_{q_u= u,c,t}\left(V^\ast_{q_u b} V^{}_{q_u s}\right) p^{q_u}_f \; .
\end{equation}
(The distinction between tree and penguin contributions is a heuristic
one, the separation by the operator that enters is more precise. For a
detailed discussion of the more complete operator product approach,
which also includes higher order QCD corrections, see, for example,
\Bref{Buchalla:1995vs}.)  Using CKM unitarity, these decay
amplitudes can always be written in terms of just two CKM
combinations:
\begin{equation}\label{btoccs}
A_{\psi K}=\left(V^\ast_{cb} V^{}_{cs}\right)T_{\psi
  K}+\left(V^\ast_{ub} V^{}_{us}\right)P^u_{\psi K},
\end{equation}
where $T_{\psi K}=t_{\psi K}+p^c_{\psi K}-p^t_{\psi K}$ and $P^u_{\psi
K}=p^u_{\psi K}-p^t_{\psi K}$. A subtlety arises in this decay that is
related to the fact that ${B}^0\to J/\psi K^0$ and $\overline{B}^0\to
J/\psi\overline{K}{}^0$. A common final state, \eg $J/\psi K_S$, can
be reached via $K^0$--$\overline{K}{}^0$ mixing.  Consequently, the
phase factor corresponding to neutral $K$ mixing,
$e^{-i\phi_K}=(V^*_{cd}V^{}_{cs})/(V^{}_{cd}V^*_{cs})$, plays a role:
\begin{equation}\label{psikmix}
\frac{\overline{A}_{\psi K_S}}{A_{\psi K_S}}
=-\frac{\left(V^{}_{cb} V^\ast_{cs}\right)T_{\psi
    K}+\left(V^{}_{ub} V^\ast_{us}\right)P^u_{\psi K}}
{\left(V^\ast_{cb} V^{}_{cs}\right)T_{\psi
    K}+\left(V^\ast_{ub} V^{}_{us}\right)P^u_{\psi K}}\times
\frac{V_{cd}^\ast V_{cs}^{}}{V_{cd}^{}V_{cs}^\ast}.
\end{equation}

The crucial point is that, for $B\to J/\psi K_S$ and other $\bar
b\to\bar cc\bar s$ processes, we can neglect the $P^u$ contribution to
$A_{\psi K}$, in the SM, to an approximation that is better than one
per cent:
\begin{equation}\label{smapprox}
|P^u_{\psi K}/T_{\psi K}|\times|V_{ub}/V_{cb}|\times|
V_{us}/V_{cs}|\sim(\text{loop\ factor})\times0.1\times0.23\lesssim0.005.
\end{equation}
Thus, to an accuracy of better than one per cent,
\begin{equation}
\lambda_{\psi K_S}=\left(\frac{V_{tb}^*
  V_{td}^{}}{V_{tb}^{}V_{td}^*}\right)\left(\frac{V_{cb}
  V_{cd}^{*}}{V_{cb}^{*}V_{cd}}\right)=-e^{-2i\beta},
\end{equation}
where $\beta$ is defined in \Eref{abcangles}, and consequently
\begin{equation}\label{btopsik}
S_{\psi K_S}=\sin2\beta,\ \ \ C_{\psi K_S}=0 \; .
\end{equation}
(Below the per cent level, several effects modify this equation
\cite{Grossman:2002bu,Boos:2004xp,Li:2006vq,Gronau:2008cc}.)

\textbf{Exercise 3:} \emph{Show that, if the $B\to\pi\pi$ decays were dominated
by tree diagrams, then $S_{\pi\pi}=\sin2\alpha$.}

\textbf{Exercise 4:} \emph{Estimate the accuracy of the predictions
$S_{\phi K_S}=\sin2\beta$ and $C_{\phi K_S}=0$.}

When we consider extensions of the SM, we still do not expect any
significant new contribution to the tree level decay, $b\to c\bar cs$,
beyond the SM $W$-mediated diagram. Thus the expression $\bar A_{\psi
K_S}/A_{\psi K_S}=(V_{cb}V_{cd}^*)/(V_{cb}^*V_{cd})$ remains valid,
though the approximation of neglecting sub-dominant phases can be
somewhat less accurate than \Eref{smapprox}. On the other hand,
$M_{12}$, the $B^0$--$\overline{B}^0$ mixing amplitude, can in
principle get large and even dominant contributions from new
physics. We can parametrize the modification to the SM in terms of two
parameters, $r_d^2$ signifying the change in magnitude, and
$2\theta_d$ signifying the change in phase:
\begin{equation}\label{derthed}
M_{12}=r_d^2\ e^{2i\theta_d}\ M_{12}^\text{SM}(\rho,\eta).
\end{equation}
This leads to the following generalization of \Eref{btopsik}:
\begin{equation}\label{btopsiknp}
S_{\psi K_S}=\sin(2\beta+2\theta_d),\ \ \ C_{\psi K_S}=0 \; .
\end{equation}

The experimental measurements give the following ranges \cite{hfag}:
\begin{equation}\label{scpkexp}
S_{\psi K_S}=0.671\pm0.024,\ \ \ C_{\psi K_S}=0.005\pm0.019 \; .
\end{equation}

\subsection{Self-consistency of the CKM assumption}

The three-generation Standard Model has room for CP violation, through
the KM phase in the quark mixing matrix. Yet, one would like to make
sure that CP is indeed violated by the SM interactions, namely that
$\sin\delta_\text{KM}\neq0$. If we establish that this is the case, we
would further like to know whether the SM contributions to CP
violating observables are dominant. More quantitatively, we would like
to put an upper bound on the ratio between the new physics and the SM
contributions.

\begin{figure}[tb]
  \centering
  \includegraphics[width=0.65\linewidth]{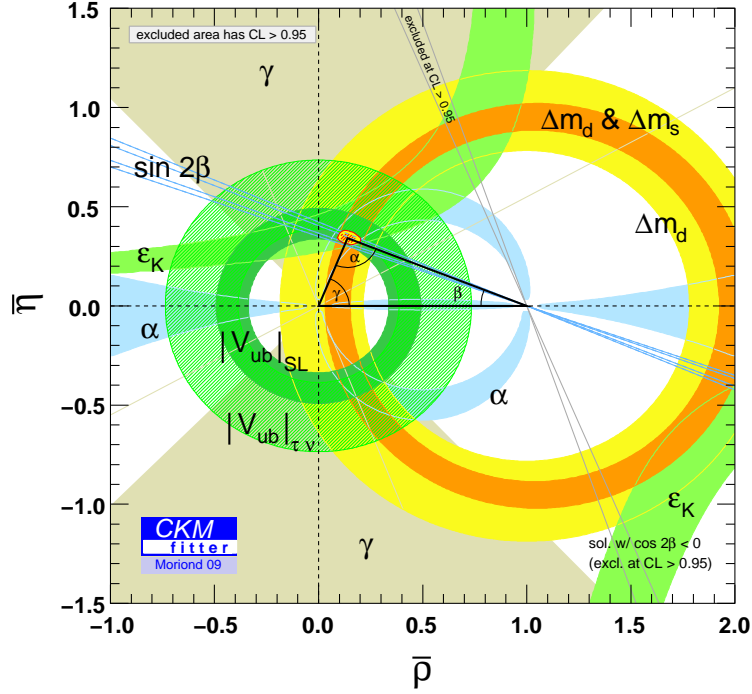}
  \caption[]{Allowed region in the $\rho$--$\eta$ plane. Superimposed are
             the individual constraints from charmless semileptonic
             $B$ decays ($|V_{ub}/V_{cb}|$), mass differences in the
             $B^0$ ($\Delta m_d$) and $B_s$ ($\Delta m_s$) neutral
             meson systems, and CP violation in $K\to\pi\pi$
             ($\varepsilon_K$), $B\to\psi K$ ($\sin2\beta$),
             $B\to\pi\pi,\rho\pi,\rho\rho$ ($\alpha$), and $B\to DK$
             ($\gamma$). Taken from \Bref{ckmfitter}.}
  \label{fg:UT}
\end{figure}

As a first step, one can assume that flavour-changing processes are
fully described by the SM, and check the consistency of the various
measurements with this assumption. There are four relevant mixing
parameters, which can be taken to be the Wolfenstein parameters
$\lambda$, $A$, $\rho$, and $\eta$ defined in \Eref{wolpar}. The values
of $\lambda$ and $A$ are known rather accurately \cite{Yao:2006px}
from, respectively, $K\to\pi\ell\nu$ and $b\to c\ell\nu$ decays:
\begin{equation}\label{lamaexp}
\lambda=0.2257\pm0.0010,\ \ \ A=0.814\pm0.022.
\end{equation}
Then, one can express all the relevant observables as a function of
the two remaining parameters, $\rho$ and $\eta$, and check whether
there is a range in the $\rho$--$\eta$ plane that is consistent with
all measurements. The list of observables includes the following:
\begin{itemize}
\item The rates of inclusive and exclusive charmless semileptonic $B$
      decays depend on $|V_{ub}|^2\propto\rho^2+\eta^2$.
\item The CP asymmetry in $B\to\psi K_S$, $S_{\psi
      K_S}=\sin2\beta=\frac{2\eta(1-\rho)}{(1-\rho)^2+\eta^2}$.
\item The rates of various $B\to DK$ decays depend on the phase
      $\gamma$, where $e^{i\gamma}=\frac{\rho+i\eta}{\rho^2+\eta^2}$.
\item The rates of various $B\to\pi\pi,\rho\pi,\rho\rho$ decays depend
      on the phase $\alpha=\pi-\beta-\gamma$.
\item The ratio between the mass splittings in the neutral $B$ and
      $B_s$ systems is sensitive to
      $|V_{td}/V_{ts}|^2=\lambda^2[(1-\rho)^2+\eta^2]$.
\item The CP violation in $K\to\pi\pi$ decays, $\epsilon_K$, depends
      in a complicated way on $\rho$ and $\eta$.
\end{itemize}
The resulting constraints are shown in \Fref{fg:UT}.

The consistency of the various constraints is impressive. In
particular, the following ranges for $\rho$ and $\eta$ can account for
all the measurements \cite{Yao:2006px}:
\begin{equation}
\rho=0.135^{+0.031}_{-0.016},\ \ \ \eta=0.349\pm0.017.
\end{equation}

One can then make the following statement \cite{Nir:2002gu}:\\
\textbf{Very likely, CP violation in flavour-changing processes is
        dominated by the Kobayashi--Maskawa phase.}

In the next two subsections, we explain how we can remove the phrase
`very likely' from this statement, and how we can quantify the
KM~dominance.

\subsection{Is the Kobayashi--Maskawa mechanism at work?}

In proving that the KM mechanism is at work, we assume that
charged-current tree-level processes are dominated by the $W$-mediated
SM diagrams (see, for example, \Bref{Grossman:1997dd}).  This is a
very plausible assumption. I am not aware of any viable well-motivated
model where this assumption is not valid.  Thus we can use all tree-level 
processes and fit them to $\rho$ and $\eta$, as we did
before. The list of such processes includes the following:
\begin{enumerate}
\item Charmless semileptonic $B$-decays, $b\to u\ell\nu$, measure
      $R_u$ [see \Eref{RbRt}].
\item $B\to DK$ decays, which go through the quark transitions $b\to
      c\bar u s$ and $b\to u\bar cs$, measure the angle $\gamma$ [see
      \Eref{abcangles}].
\item $B\to\rho\rho$ decays (and, similarly, $B\to\pi\pi$ and
      $B\to\rho\pi$ decays) go through the quark transition $b\to
      u\bar ud$. With an isospin analysis, one can determine the
      relative phase between the tree decay amplitude and the mixing
      amplitude. By incorporating the measurement of $S_{\psi K_S}$,
      one can subtract the phase from the mixing amplitude, finally
      providing a measurement of the angle $\gamma$ [see
      \Eref{abcangles}].
  \end{enumerate}

In addition, we can use loop processes, but then we must allow for new
physics contributions, in addition to the $(\rho,\eta)$-dependent SM
contributions. Of course, if each such measurement adds a separate
mode-dependent parameter, then we do not gain anything by using this
information. However, there are a number of observables where the only
relevant loop process is $B^0$--$\overline{B}{}^0$ mixing. The list
includes $S_{\psi K_S}$, $\Delta m_B$, and the CP asymmetry in
semileptonic $B$ decays:
\begin{align}\label{apksNP}
S_{\psi K_S}         &=\sin(2\beta+2\theta_d),\nonumber\\
\Delta m_{B}         &=r_d^2(\Delta m_B)^\text{SM},\nonumber\\
\mathcal{A}_\text{SL}&=-
  \mathcal{R}e \left(\frac{\Gamma_{12}}{M_{12}}\right)^\text{SM}
               \frac{\sin2\theta_d}{r_d^2}
  +\mathcal{I}m\left(\frac{\Gamma_{12}}{M_{12}}\right)^\text{SM}
               \frac{\cos2\theta_d}{r_d^2}.
\end{align}
As explained above, such processes involve two new parameters [see
\Eref{derthed}]. Since there are three relevant observables, we can
further tighten the constraints in the $(\rho,\eta)$~plane. Similarly,
one can use measurements related to $B_s$--$\overline{B}_s$
mixing. One gains three new observables at the cost of two new
parameters (see, for example, \Bref{Grossman:2006ce}).

\begin{figure}[tb]
  \centering
  \includegraphics[width=0.65\linewidth]{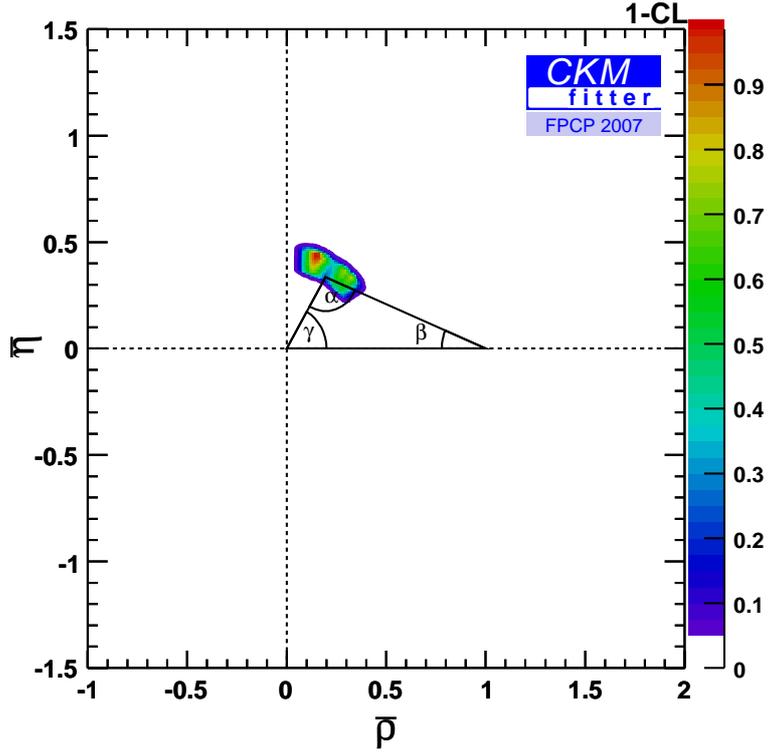}
  \caption[]{The allowed region in the $\rho$--$\eta$ plane, assuming
             that tree diagrams are dominated by the Standard Model
             \cite{ckmfitter}}
  \label{fig:re_tree}
\end{figure}

The results of such a fit, projected on the $\rho$--$\eta$ plane, can be
seen in \Fref{fig:re_tree}. It gives \cite{ckmfitter}
\begin{equation}
\eta=0.44^{+0.05}_{-0.23}\ \ (3\sigma).
\end{equation}
[A similar analysis in \Bref{Bona:2007vi} obtains the $3\sigma$
range $(0.31$--$0.46)$.] It is clear that $\eta\neq0$ is well
established:\\
\textbf{The Kobayashi--Maskawa mechanism of CP violation is at work.}

Another way to establish that CP is violated by the CKM matrix is to
find, within the same procedure, the allowed range for $\sin2\beta$
\cite{Bona:2007vi}:
\begin{equation}\label{stbth}
\sin2\beta^\text{tree}=0.76\pm0.04.
\end{equation}
(\Bref[b]{ckmfitter} finds $0.82^{+0.02}_{-0.13}$.) Thus,
$\beta\neq0$ is well established.

The consistency of the experimental results (\ref{scpkexp}) with the
SM predictions (\ref{btopsik},\ref{stbth}) means that the KM mechanism
of CP violation dominates the observed CP violation. In the next
subsection, we make this statement more quantitative.

\subsection{How much can new physics contribute to
            $B^0$--$\overline{B}{}^0$ mixing?}

\begin{figure}[htb]
\centering
\raisebox{60mm}{(a)}%
\includegraphics[width=.45\linewidth]{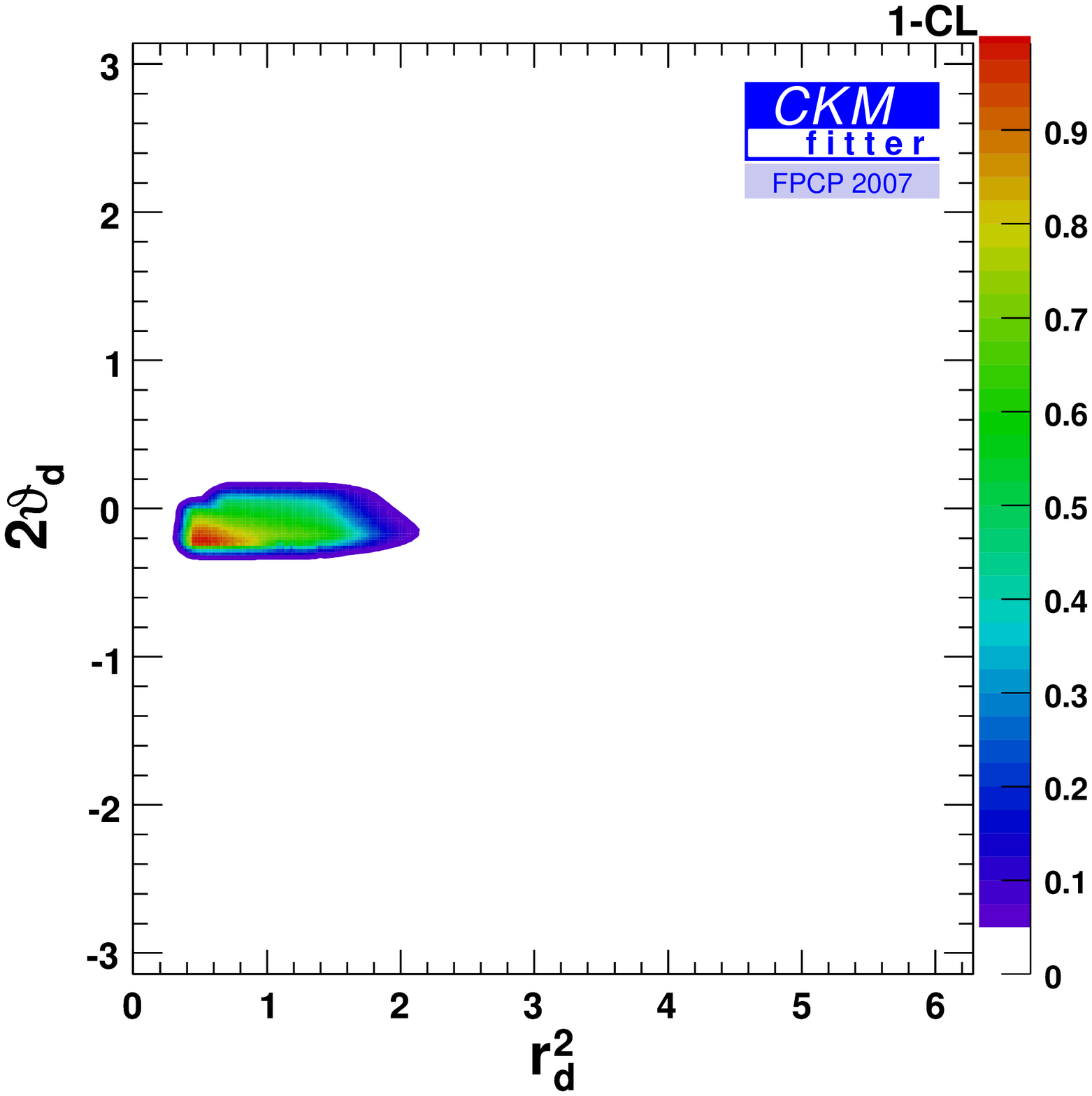}
\quad
\raisebox{60mm}{(b)}%
\includegraphics[width=.45\linewidth]{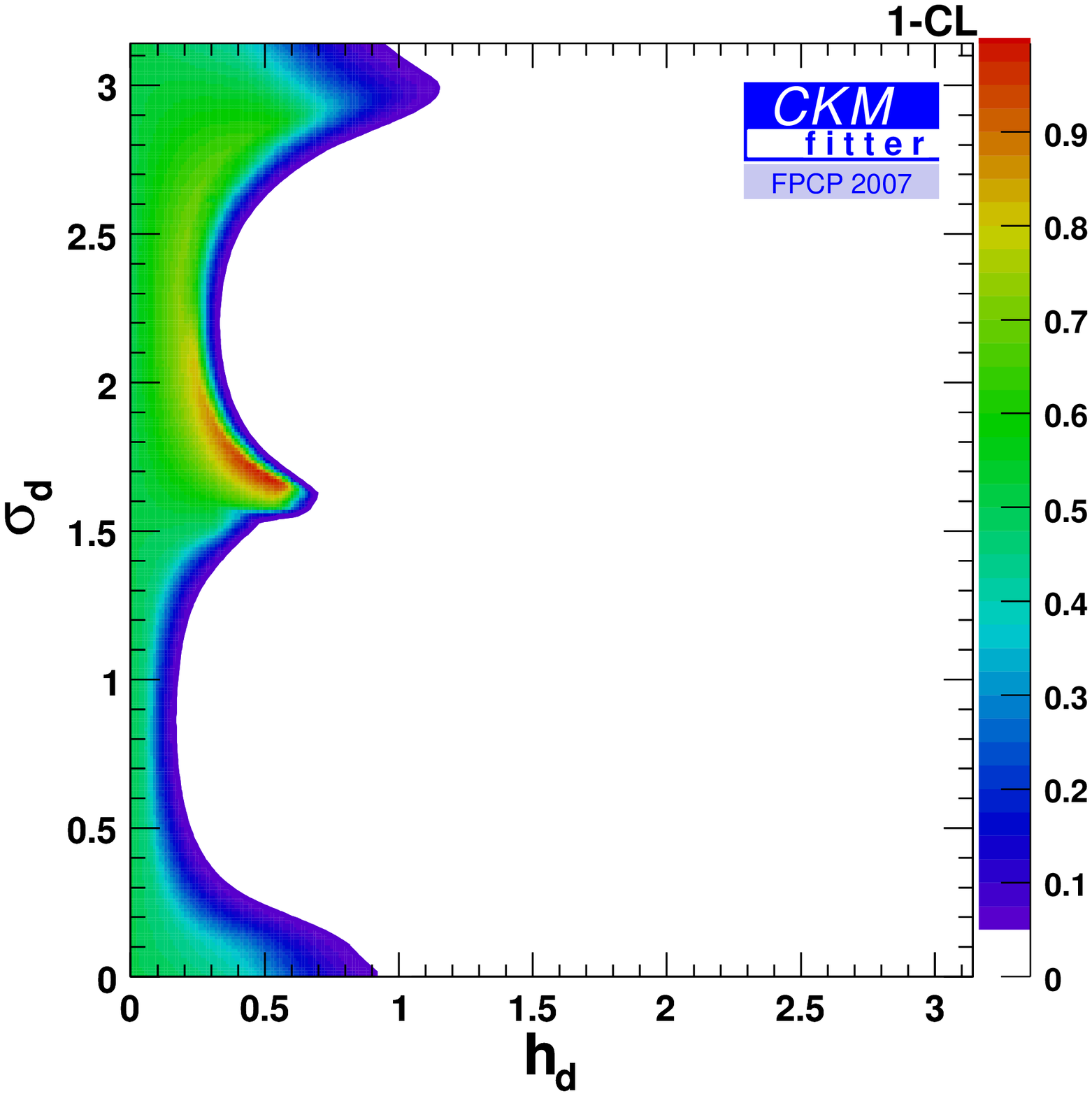}
\caption[]{Constraints in the (a) $r_d^2$--$2\theta_d$ plane, and (b)
           $h_d$--$\sigma_d$ plane, assuming that new physics
           contributions to tree-level processes are negligible
           \cite{ckmfitter}}
\label{fig:rdtd}
\end{figure}

All that we need to do in order to establish whether the SM dominates
the observed CP violation, and to put an upper bound on the new
physics contribution to $B^0$--$\overline{B}{}^0$ mixing, is to
project the results of the fit performed in the previous subsection on
the $r_d^2$--$2\theta_d$ plane. If we find that $\theta_d\ll\beta$, then
the SM dominance in the observed CP violation will be established.
The constraints are shown in \Fref{fig:rdtd}(a). Indeed,
$\theta_d\ll\beta$.

An alternative way to present the data is to use the $h_d,\sigma_d$
parametrization,
\begin{equation}
r_d^2e^{2i\theta_d}=1+h_d e^{2i\sigma_d}.
\end{equation}
While the $r_d,\theta_d$ parameters give the relation between the full
mixing amplitude and the SM one, and are convenient to apply to the
measurements, the $h_d,\sigma_d$ parameters give the relation between
the new physics and SM contributions, and are more convenient in
testing theoretical models:
\begin{equation}
h_de^{i\sigma_d}=\frac{M_{12}^\text{NP}}{M_{12}^\text{SM}}.
\end{equation}
The constraints in the $h_d$--$\sigma_d$ plane are shown in
\Fref{fig:rdtd}(b). We can make the following two statements:
\begin{enumerate}
\item A new physics contribution to the $B^0$--$\overline{B}^0$ mixing
      amplitude that carries a phase that is significantly different
      from the KM phase is constrained to lie below the $20$--$30$\%
      level.
\item A new physics contribution to the $B^0$--$\overline{B}^0$ mixing
      amplitude which is aligned with the KM phase is constrained to
      be at most comparable to the CKM contribution.
\end{enumerate}
One can reformulate these statements as follows:
\begin{enumerate}
\item The KM mechanism dominates CP violation in
      $B^0$--$\overline{B}^0$ mixing.
\item The CKM mechanism is a major player in $B^0$--$\overline{B}^0$
      mixing.
\end{enumerate}

\section{The new physics flavour puzzle}
\label{sec:nppuzzle}

It is clear that the Standard Model is not a complete theory of
Nature:
\begin{enumerate}
\item It does not include gravity, and therefore it cannot be valid at
      energy scales above $m_\text{Planck}\sim10^{19}\UGeV$.
\item It does not allow for neutrino masses, and therefore it cannot
      be valid at energy scales above
      $m_\text{seesaw}\sim10^{15}\UGeV$.
\item The fine-tuning problem of the Higgs mass and the puzzle of dark
      matter suggest that the scale where the SM is replaced with a
      more fundamental theory is actually much lower,
      $\Lambda_\text{NP}\lesssim1\UTeV$.
\end{enumerate}
Given that the SM is only an effective low-energy theory,
non-renormalizable terms must be added to $\mathcal{L}_\text{SM}$ of
\Eref{LagSM}. These are terms of dimension higher than four in the
fields which, therefore, have couplings that are inversely
proportional to the scale of new physics $\Lambda_\text{NP}$. For
example, the lowest-dimension non-renormalizable terms are dimension
five:
\begin{equation}\label{Hnint}
-\mathcal{L}_\text{Yukawa}^\text{dim$-5$}=
\frac{Z_{ij}^\nu}{\Lambda_\text{NP}}L_{Li}^I L_{Lj}^I\phi\phi+\text{h.c.}
\end{equation}
These are the seesaw terms, leading to neutrino masses. We shall
return to the topic of neutrino masses in \Sref{sec:nu}.

\textbf{Exercise 5:} \emph{How does the global symmetry breaking pattern
(\ref{globre}) change when (\ref{Hnint}) is taken into account?}

\textbf{Exercise 6:} \emph{What is the number of physical lepton flavour
parameters in this case? Identify these parameters in the mass basis.}

As concerns quark flavour physics, consider, for example, the following
dimension-six, four-fermion, flavour-changing operators:
\begin{equation}\label{eq:ffll}
\mathcal{L}_{\Delta F=2}=
 \frac{z_{sd}}{\Lambda_\text{NP}^2}(\overline{d_L}\gamma_\mu s_L)^2
+\frac{z_{cu}}{\Lambda_\text{NP}^2}(\overline{c_L}\gamma_\mu u_L)^2
+\frac{z_{bd}}{\Lambda_\text{NP}^2}(\overline{d_L}\gamma_\mu b_L)^2
+\frac{z_{bs}}{\Lambda_\text{NP}^2}(\overline{s_L}\gamma_\mu b_L)^2.
\end{equation}
Each of these terms contributes to the mass splitting between the
corresponding two neutral mesons. For example, the term
$\mathcal{L}_{\Delta B=2}\propto(\overline{d_L}\gamma_\mu b_L)^2$
contributes to $\Delta m_B$, the mass difference between the two
neutral $B$-mesons. We use $M_{12}^B=\frac{1}{2m_B}\langle
B^0|\mathcal{L}_{\Delta F=2}|\overline{B}^0\rangle$ and
\begin{equation}
\langle B^0|(\overline{d_{La}}\gamma^\mu
b_{La})(\overline{d_{Lb}}\gamma_\mu b_{Lb})|\overline{B}^0\rangle =
-\frac13 m_B^2f_B^2 B_B.
\end{equation}
Analogous expressions hold for the other neutral mesons\footnote
   {The PDG \cite{Yao:2006px} quotes the following values, extracted
    from leptonic charged meson decays: $f_K\approx0.16\UGeV$,
    $f_D\approx0.23\UGeV$, $f_B\approx0.18\UGeV$. We further use
    $f_{B_s}\approx0.20\UGeV$.}.
This leads to $\Delta
m_B/m_B=2|M_{12}^B|/m_B\sim (|z_{bd}|/3)(f_B/\Lambda_\text{NP})^2$.
Experiments give, for CP conserving observables (the experimental
evidence for $\Delta m_D$ is at the $3\sigma$ level):
\begin{eqnarray}
\Delta m_K/m_K&\sim&7.0\times10^{-15},\nonumber\\
\Delta m_D/m_D&\sim&8.7\times10^{-15},\nonumber\\
\Delta m_B/m_B&\sim&6.3\times10^{-14},\nonumber\\
\Delta m_{B_s}/m_{B_s}&\sim&2.1\times10^{-12},
\end{eqnarray}
and for CP violating ones
\begin{eqnarray}
\epsilon_K&\sim&2.3\times10^{-3},\nonumber\\
A_\Gamma/y_{\rm CP}&\lsim&0.2,\nonumber\\
S_{\psi K_S}&=&0.67\pm0.02,\nonumber\\
S_{\psi\phi}&\lsim&1.
\end{eqnarray}
These measurements give then the following constraints:
\begin{equation}
\label{lowlnp1}
\Lambda_{\rm NP}\gsim
\begin{cases}
  \sqrt{z_{sd}}\ 1\times10^3\ \textrm{TeV}&\Delta m_K\\
  \sqrt{z_{cu}}\ 1\times10^3\ \textrm{TeV}&\Delta m_D\\
  \sqrt{z_{bd}}\ 4\times10^2\ \textrm{TeV}&\Delta m_B\\
  \sqrt{z_{bs}}\ 7\times10^1\ \textrm{TeV}&\Delta m_{B_s}
\end{cases}
\end{equation}
and, for maximal phases,
\begin{equation}
\label{lowlnp2}
\Lambda_{\rm NP}\gsim
\begin{cases}
  \sqrt{z_{sd}}\ 2\times10^4\ \textrm{TeV}&\epsilon_K\\
  \sqrt{z_{cu}}\ 3\times10^3\ \textrm{TeV}&A_\Gamma\\
  \sqrt{z_{bd}}\ 8\times10^2\ \textrm{TeV}&S_{\psi K}\\
  \sqrt{z_{bs}}\ 7\times10^1\ \textrm{TeV}&S_{\psi\phi}
\end{cases}
\end{equation}
If the new physics has a generic flavour structure, that is
$z_{ij}={\cal O}(1)$, then its scale must be above $10^3$--$10^4$~TeV
(or, if the leading contributions involve electroweak loops, above
$10^2$--$10^3$~TeV).\footnote{The bounds from the corresponding four-fermi
 terms with LR structure, instead of the LL structure of
 Eq. (\ref{eq:ffll}), are even stronger.}

 {\it If indeed $\Lambda_{\rm NP}\gg \textrm{TeV}$, it means
that we have misinterpreted the hints from the fine-tuning problem
and the dark matter puzzle.} There is, however, another way to look
at these constraints:
\begin{eqnarray}
\label{zcons1}
z_{sd}&\lsim&8\times10^{-7}\ (\Lambda_{\rm NP}/\textrm{TeV})^2,\nonumber\\
z_{cu}&\lsim&5\times10^{-7}\ (\Lambda_{\rm NP}/\textrm{TeV})^2,\nonumber\\
z_{bd}&\lsim&5\times10^{-6}\ (\Lambda_{\rm NP}/\textrm{TeV})^2,\nonumber\\
z_{bs}&\lsim&2\times10^{-4}\ (\Lambda_{\rm NP}/\textrm{TeV})^2,
\end{eqnarray}
\begin{eqnarray}
\label{zcons2}
z_{sd}^I&\lsim&6\times10^{-9}\ (\Lambda_{\rm NP}/\textrm{TeV})^2,\nonumber\\
z_{cu}^I&\lsim&1\times10^{-7}\ (\Lambda_{\rm NP}/\textrm{TeV})^2,\nonumber\\
z_{bd}^I&\lsim&1\times10^{-6}\ (\Lambda_{\rm NP}/\textrm{TeV})^2,\nonumber\\
z_{bs}^I&\lsim&2\times10^{-4}\ (\Lambda_{\rm NP}/\textrm{TeV})^2.
\end{eqnarray}
{\it It could be that the scale of new physics is of order TeV, but
  its flavour structure is far from generic.}

One can use that language of effective operators also for the SM,
integrating out all particles significantly heavier than the neutral
mesons (that is, the top, the Higgs, and the weak gauge bosons). Thus
the scale is $\Lambda_\text{SM}\sim m_W$. Since the leading
contributions to neutral meson mixings come from box diagrams, the
$z_{ij}$ coefficients are suppressed by $\alpha_2^2$. To identify the
relevant flavour suppression factor, one can employ the spurion
formalism. For example, the flavour transition that is relevant to
$B^0$--$\overline{B}{}^0$ mixing involves $\overline{d_L}b_L$ which
transforms as $(8,1,1)_{SU(3)_q^3}$. The leading contribution must
then be proportional to $(Y^u Y^{u\dagger})_{13}\propto y_t^2
V_{tb}V_{td}^*$. Indeed, an explicit calculation (using VIA for the
matrix element and neglecting QCD corrections) gives\footnote
     {A detailed derivation can be found in Appendix B of \Bref{Branco:1999fs}.}
\begin{equation} \frac{2M_{12}^B}{m_B}\approx-\frac{\alpha_2^2}{12}
\frac{f_B^2}{m_W^2}S_0(x_t)(V_{tb}V_{td}^*)^2,
\end{equation}
where $x_i=m_i^2/m_W^2$ and
\begin{equation}
S_0(x)=\frac{x}{(1-x)^2}\left[1-\frac{11x}{4}+\frac{x^2}{4}-\frac{3x^2\ln
x}{2(1-x)}\right].  \end{equation}
Similar spurion analyses, or explicit calculations, allow us to
extract the weak and flavour suppression factors that apply in the
SM:
\begin{align}
\mathcal{I}m(z_{sd}^\text{SM})
  &\sim \alpha_2^2 y_t^2 |V_{td}V_{ts}|^2\sim1\times10^{-10},\nonumber\\
z_{sd}^\text{SM}
  &\sim \alpha_2^2 y_c^2 |V_{cd}V_{cs}|^2\sim5\times10^{-9},\nonumber\\
z_{bd}^\text{SM}
  &\sim \alpha_2^2 y_t^2 |V_{td}V_{tb}|^2\sim7\times10^{-8},\nonumber\\
z_{bs}^\text{SM}
  &\sim \alpha_2^2 y_t^2 |V_{ts}V_{tb}|^2\sim2\times10^{-6}.
\end{align}
(We did not include $z_{cu}^\text{SM}$ in the list because it requires
a more detailed consideration. The naively leading short distance
contribution is $\propto \alpha_2^2(y_s^4/y_c^2)
|V_{cs}V_{us}|^2\sim5\times10^{-13}$. However, higher dimension terms
can replace a $y_s^2$ factor with $(\Lambda/m_D)^2$
\cite{Bigi:2000wn}. Moreover, long distance contributions are expected
to dominate. In particular, peculiar phase space effects
\cite{Falk:2001hx,Falk:2004wg} have been identified which are expected
to enhance $\Delta m_D$ to within an order of magnitude of its
measured value.)

It is clear then that contributions from new physics at
$\Lambda_\text{NP}\sim1\UTeV$ should be suppressed by factors that are
comparable to or smaller than the SM ones. Why does that happen? This
is the new physics flavour puzzle.

The fact that the flavour structure of new physics at the \UTeVZ{} scale
must be non-generic means that flavour measurements are a good probe of
the new physics. Perhaps the best-studied example is that of
supersymmetry. Here, the spectrum of the superpartners and the
structure of their couplings to the SM fermions will allow us to probe
the mechanism of dynamical supersymmetry breaking.

\section{Lessons for supersymmetry from $D^0$--$\overline{D}^0$ mixing}
\label{sec:dmix}

Interesting experimental results concerning $D^0$--$\overline{D}^0$
mixing have recently been achieved by the BELLE and BaBar experiments.
For the first time, there is evidence for width splitting
\cite{Aubert:2007wf,Staric:2007dt} and mass splitting (of order one
per cent) between the two neutral $D$-mesons. Allowing for indirect CP
violation, the world averages of the mixing parameters are \cite{hfag}
\begin{align}
x&=(1.00\pm0.25)\times10^{-2},\nonumber\\
y&=(0.77\pm0.18)\times10^{-2}.
\end{align}
It is important to note, however, that there is no evidence for CP
violation in this mixing \cite{hfag}:
\begin{align}\label{eq:cpvd}
1-|q/p|&=+0.06\pm0.14,\nonumber\\
\phi_D &=-0.04\pm0.09.
\end{align}
We use this recent experimental
information to draw important lessons on supersymmetry. This
demonstrates how flavour physics---at the \UGeVZ{} scale---provides a
significant probe of supersymmetry---at the \UTeVZ{} scale.

\subsection{Neutral meson mixing with supersymmetry}
We consider the contributions from the box diagrams involving the
squark doublets of the first two generations, $\tilde Q_{L1,2}$, to
the $D^0$--$\overline{D}^0$ and $K^0$--$\overline{K}^0$ mixing
amplitudes.  The contributions that are relevant to the neutral $D$
system are proportional to $K_{2i}^u K^{u*}_{1i}K_{2j}^u K^{u*}_{1j}$,
where $K^u$ is the mixing matrix of the gluino couplings to a
left-handed up quark and their supersymmetric squark partners. (In the
language of the mass insertion approximation, we calculate here the
contribution that is $\propto[(\delta^u_{LL})_{12}]^2$.) The
contributions that are relevant to the neutral $K$ system are
proportional to $K_{2i}^{d*} K^{d}_{1i}K_{2j}^{d*} K^{d}_{1j}$, where
$K^d$ is the mixing matrix of the gluino couplings to a left-handed
down quark and their supersymmetric squark partners
($\propto[(\delta^d_{LL})_{12}]^2$ in the mass insertion
approximation). We work in the mass basis for both quarks and squarks.
A detailed derivation \cite{Raz:2002zx} is given in Appendix
\ref{app:susyd}. It gives
\begin{align}\label{motsusyb}
M_{12}^D
  &=\frac{\alpha_s^2m_Df_D^2B_D\eta_\text{QCD}}{108m_{\tilde u}^2}
    [11\tilde f_6(x_u)+4x_uf_6(x_u)]\frac{(\Delta m^2_{\tilde
    u})^2}{m_{\tilde u}^4} (K_{21}^uK_{11}^{u*})^2,\\
\label{motsusyc}
M_{12}^K
  &=\frac{\alpha_s^2m_Kf_K^2B_K\eta_\text{QCD}}{108m_{\tilde d}^2}
    [11\tilde f_6(x_d)+4x_df_6(x_d)]\frac{(\Delta\tilde m^2_{\tilde
    d})^2}{\tilde m_d^4} (K_{21}^{d*}K_{11}^{d})^2.
\end{align}
Here $m_{\tilde u,\tilde d}$ is the average mass of the corresponding
two squark generations, $\Delta m^2_{\tilde u,\tilde d}$ is the
mass-squared difference, and $x_{u,d}=m_{\tilde g}^2/m_{\tilde
  u,\tilde d}^2$.

One can immediately identify three generic ways in which
supersymmetric contributions to neutral meson mixing can be
suppressed:
\begin{enumerate}
\item Heaviness: $m_{\tilde q}\gg1\UTeV$.
\item Degeneracy: $\Delta m^2_{\tilde q}\ll m_{\tilde q}^2$.
\item Alignment: $K^{d,u}_{21}\ll1$.
\end{enumerate}
When heaviness is the only suppression mechanism, as in split
supersymmetry \cite{ArkaniHamed:2004fb}, the squarks are very heavy
and supersymmetry no longer solves the fine tuning problem\footnote
    {When the first two squark generations are mildly heavy and the
     third generation is light, as in effective supersymmetry
     \cite{Cohen:1996vb}, the fine tuning problem is still solved, but
     additional suppression mechanisms are needed.}.
If we want to maintain supersymmetry as a solution to the fine tuning
problem, either degeneracy, or alignment, or a combination of both is
needed. This means that the flavour structure of supersymmetry is not
generic, as argued in the previous section.

The $2\times2$ mass-squared matrices for the relevant squarks have the
following form:
\begin{align}\label{mllot}
\tilde M^2_{U_L}
  &= \tilde m^2_{Q_L}
    +\left(\frac{1}{2}-\frac{2}{3}s^2_W\right)m_Z^2\cos2\beta
    +M_u M_u^\dagger,\nonumber\\
\tilde M^2_{D_L}
  &= \tilde m^2_{Q_L}
    -\left(\frac{1}{2}-\frac{1}{3}s^2_W\right)m_Z^2\cos2\beta
    +M_d M_d^\dagger.
\end{align}
We note the following features of the various terms:
\begin{itemize}
\item $\tilde m^2_{Q_L}$ is a $2\times2$ Hermitian matrix of soft
      supersymmetry breaking terms. It does not break $SU(2)_\text{L}$
      and consequently it is common to $\tilde M^2_{U_L}$ and $\tilde
      M^2_{D_L}$. On the other hand, it breaks in general the
      $SU(2)_Q$ flavour symmetry.
\item The terms proportional to $m_Z^2$ are the D~terms. They break
      supersymmetry (since they involve $D_{T_3}\neq0$ and $D_Y\neq0$)
      and $SU(2)_\text{L}$ but conserve $SU(2)_Q$.
\item The terms proportional to $M_q^2$ come from the $F_{U_R}$ and
      $F_{D_R}$ terms. They break the gauge $SU(2)_\text{L}$ and the
      global $SU(2)_Q$ but, since $F_{U_R}=F_{D_R}=0$, conserve
      supersymmetry.
\end{itemize}
Given that we are interested in squark masses close to the \UTeVZ{} scale
(and the experimental lower bounds are of order $300\UGeV$), the scale of the
eigenvalues of $\tilde m^2_{Q_L}$ is much higher than
$m_Z^2$ which, in turn, is much higher than $m_c^2$, the largest
eigenvalue in $M_q M_q^\dagger$ (in the two-generation framework). We can draw the following conclusions:
\begin{enumerate}
\item $m_{\tilde u}^2=m_{\tilde d}^2\equiv m_{\tilde q}^2$ up to
      effects of order $m_Z^2$, namely to an accuracy of
      $\mathcal{O}(10^{-2})$.
\item $\Delta m^2_{\tilde u}=\Delta m^2_{\tilde d}\equiv \Delta
      m^2_{\tilde q}$ up to effects of order $m_c^2$, namely to an
      accuracy of $\mathcal{O}(10^{-5})$.
\item Since $K_u\simeq V_{uL} \tilde V_L^\dagger$ and $K_d\simeq
      V_{dL} \tilde V_L^\dagger$ [the matrices $V_{qL}$ are defined in
      \Eref{diagMq}, while $\tilde V_L$ diagonalizes $\tilde
      m^2_{Q_L}$], the mixing matrices $K^u$ and $K^d$ are different
      from each other, but the following relation to the CKM matrix
      holds to an accuracy of $\mathcal{O}(10^{-5})$:
\begin{equation}\label{kkckm}
K^u K^{d\dagger} = V.
\end{equation}
\end{enumerate}

\subsection{Non-degenerate squarks at the LHC?}

\Erefs[b]{motsusyb} and (\ref{motsusyc}) can be translated into our
generic language:
\begin{align}
\Lambda_\text{NP}&= m_{\tilde q},\\
z_{cu}           &= z_{12}\sin^2\theta_u,\nonumber\\
z_{sd}           &= z_{12}\sin^2\theta_d,\nonumber\\
z_{12}           &= \frac{11\tilde f_6(x)+4x f_6(x)}{18}
                    \alpha_s^2
                    \left(\frac{\Delta\tilde m_{\tilde q}^2}{m_{\tilde q}^2}\right)^2,
\end{align}
with \Eref{kkckm} giving
\begin{equation}\label{kkckmb}
\sin\theta_u-\sin\theta_d\approx\sin\theta_c=0.23.
\end{equation}

We now ask the following question: Is it possible that the first 
two-generation squarks, $\tilde Q_{L1,2}$, are accessible to the LHC
($m_{\tilde q}\lesssim1\UTeV$), and are not degenerate ($\Delta
m^2_{\tilde q}/m_{\tilde q}^2=\mathcal{O}(1)$)?

To answer this question, we use Eqs.~(\ref{zcons1}) and (\ref{zcons2}). For
$\Lambda_\text{NP}\lesssim1\UTeV$, we have
$z_{cu}\lesssim5\times10^{-7}$ and, for a phase that is $\not\ll0.1$,
$z_{sd}\lesssim6\times10^{-8}$. On the other hand, for non-degenerate
squarks, and, for example, $11\tilde f_6(1)+4f_6(1)=1/6$, we have
$z_{12}=8\times10^{-5}$. Then we need, simultaneously,
$\sin\theta_u\lesssim0.08$ and $\sin\theta_d\lesssim0.03$, but this is
inconsistent with \Eref{kkckmb}.

There are three ways out of this situation:
\begin{enumerate}
\item The first two generation squarks are quasi-degenerate. The
      minimal level of degeneracy is $(\tilde m_2-\tilde m_1)/(\tilde
      m_2+\tilde m_1)\lesssim0.1$. It could be the result of RGE
      \cite{Nir:2002ah}. However, for maximal phases, the bound is even
     stronger, of order 0.04 \cite{Blum:2009sk}, which is difficult to 
     achieve with just RGE effects. 
\item The first two generation squarks are heavy. Putting
      $\sin\theta_u=0.23$ and $\sin\theta_d\approx0$, as in models of
      alignment \cite{Nir:1993mx,Leurer:1993gy}, \Eref{lowlnp2} leads
      to
      \begin{equation}\label{mqali}
           m_{\tilde q}\gtrsim3\UTeV\SPp.
      \end{equation}
\item The ratio $x=\tilde m_g^2/\tilde m_q^2$ is in a fine-tuned
      region of parameter space where there are accidental
      cancellations in $11\tilde f_6(x)+4xf_6(x)$. For example, for
      $x=2.33$, this combination is $\sim0.003$ and the bound
      (\ref{mqali}) is relaxed by a factor of 7.
\end{enumerate}
Barring accidental cancellations, the \emph{model-independent}
conclusion is that, if the first two generations of squark doublets
are within the reach of the LHC, they must be quasi-degenerate
\cite{Ciuchini:2007cw,Nir:2007ac}. Analogous conclusions can be drawn
for many TeV-scale new physics scenarios: a strong level of degeneracy
is required (for definitions and detailed analysis, see Ref.~\cite{Blum:2009sk}).

\textbf{Exercise 7:} \emph{Does $K_{31}^d\sim|V_{ub}|$ suffice to satisfy the
$\Delta m_B$ constraint with neither degeneracy nor heaviness? (Use
the two-generation approximation and ignore the second generation.)}

Is there a natural way to make the squarks degenerate? Examining
Eqs. (\ref{mllot}) we learn that degeneracy requires $\tilde
m^2_{Q_L}\simeq\tilde m^2_{\tilde q}\mathbf{1}$. We have mentioned
already that flavour universality is a generic feature of gauge
interactions. Thus the requirement of degeneracy is perhaps a hint
that supersymmetry breaking is \emph{gauge mediated} to the MSSM
fields.

\section{Flavour at the LHC}
\label{sec:lhc}

The LHC will study the physics of electroweak symmetry breaking. There
are high hopes that it will discover not only the Higgs, but also shed
light on the fine-tuning problem that is related to the Higgs mass.
Here, we focus on the issue of how, through the study of new physics,
the LHC can shed light on the new physics flavour puzzle.

\subsection{Minimal flavour violation (MFV)}
If supersymmetry breaking is gauge mediated, the squark mass matrices
of \Eref{mllot}, and those for the SU(2)-singlet squarks, have the
following form at the scale of mediation $m_M$:
\begin{align}\label{mllgm}
\tilde M^2_{U_L}(m_M)&= \left(m^2_{\tilde Q_L}+D_{U_L}\right)
  \mathbf{1}+M_u M_u^\dagger,\nonumber\\
\tilde M^2_{D_L}(m_M)&= \left(m^2_{\tilde Q_L}+D_{D_L}\right)
  \mathbf{1}+M_d M_d^\dagger,\nonumber\\
\tilde M^2_{U_R}(m_M)&= \left(m^2_{\tilde U_R}+D_{U_R}\right)
  \mathbf{1}+M_u^\dagger M_u,\nonumber\\
\tilde M^2_{D_R}(m_M)&= \left(m^2_{\tilde D_R}+D_{D_R}\right)
  \mathbf{1}+M_d^\dagger M_d,
\end{align}
where $D_{q_A}=(T_3)_{q_A}-(Q_\text{EM})_{q_A}s^2_W m_Z^2\cos2\beta$
are the $D$-term contributions. Here, the only source of the
$SU(3)^3_q$ breaking are the SM Yukawa matrices.

This statement holds also when the renormalization group evolution is
applied to find the form of these matrices at the weak scale.  Taking
the scale of the soft breaking terms $m_{\tilde q_A}$ to be somewhat
higher than the electroweak breaking scale $m_Z$ allows us to neglect
the $D_{q_A}$ and $M_q$ terms in (\ref{mllgm}). Then we obtain
\begin{align}\label{mllrrmz}
\tilde M^2_{Q_L}(m_Z)
  &\sim m^2_{\tilde Q_L}\left(r_3\mathbf{1}+c_u
        Y_uY_u^\dagger+c_d Y_d Y_d^\dagger\right),\nonumber\\
\tilde M^2_{U_R}(m_Z)
  &\sim m^2_{\tilde U_R}\left(r_3\mathbf{1}+c_{uR}
        Y_u^\dagger Y_u\right),\nonumber\\
\tilde M^2_{D_R}(m_Z)
  &\sim m^2_{\tilde D_R}\left(r_3\mathbf{1}+c_{dR}
        Y_d^\dagger Y_d\right).
\end{align}
Here $r_3$ represent the universal RGE contribution that is
proportional to the gluino mass
($r_3=\mathcal{O}(6)\times(M_3(m_M)/m_{\tilde q}(m_M))$) and the
$c$-coefficients depend logarithmically on $m_M/m_Z$ and can be of
$\mathcal{O}(1)$ when $m_M$ is not far below the GUT scale.

Models of gauge mediated supersymmetry breaking (GMSB) provide a
concrete example of a large class of models that obey a simple
principle called \emph{minimal flavour violation} (MFV)
\cite{D'Ambrosio:2002ex}. This principle guarantees that low-energy
flavour-changing processes deviate only very little from the SM
predictions.  The basic idea can be described as follows. The gauge
interactions of the SM are universal in flavour space. The only
breaking of this flavour universality comes from the three Yukawa
matrices, $Y_U$, $Y_D$, and $Y_E$. If this remains true in the
presence of the new physics, namely $Y_U$, $Y_D$, and $Y_E$ are the
only flavour non-universal parameters, then the model belongs to the
MFV class.

Let us now formulate this principle in a more formal way, using the
language of spurions that we presented in \Sref{sec:spurions}.
The Standard Model with vanishing Yukawa couplings has a large global
symmetry of \Erefs{gglobal} and (\ref{susuu}). In this section we concentrate
only on the quarks. The non-Abelian part of the flavour symmetry for
the quarks is $SU(3)_q^3$ of \Eref{susuu} with the three generations
of quark fields transforming as follows:
\begin{equation}
Q_L(3,1,1),\ \ U_R(1,3,1),\ \ D_R(1,1,3).
\end{equation}
The Yukawa interactions,
\begin{equation}\label{eq:lagy}
\mathcal{L}_Y=\overline{Q_L}Y_D D_R H + \overline{Q_L}Y_U U_R H_c ,
\end{equation}
($H_c=i\tau_2 H^*$) break this symmetry. The Yukawa couplings can thus
be thought of as spurions with the following transformation properties
under $SU(3)_q^3$ [see \Eref{Gglobq}]:
\begin{equation}
Y_U\sim(3,\bar3,1),\qquad Y_D\sim(3,1,\bar3).
\end{equation}
When we say `spurions', we mean that we pretend that the Yukawa
matrices are fields which transform under the flavour symmetry, and
then require that all the Lagrangian terms, constructed
from the SM fields, $Y_{D}$ and $Y_U$, must be (formally)
invariant under the flavour group $SU(3)_q^3$. Of course, in reality,
$\mathcal{L}_Y$ breaks $SU(3)_q^3$ precisely because $Y_{D,U}$ are {\it
  not} fields and do not transform under the symmetry.

The idea of minimal flavour violation is relevant to extensions of the
SM, and can be applied in two ways:
\begin{enumerate}
\item If we consider the SM as a low-energy effective theory, then all
      higher-dimension operators, constructed from SM~fields and
      $Y$~spurions, are formally invariant under $G_\text{global}$.
\item If we consider a full high-energy theory that extends the SM,
      then all operators, constructed from SM and the new fields, and
      from $Y$~spurions, are formally invariant under
      $G_\text{global}$.
\end{enumerate}

\textbf{Exercise 8:} \emph{Use the spurion formalism to argue that, in MFV
models, the $K_L\to\pi^0\nu\bar\nu$ decay amplitude is proportional to
$y_t^2 V_{td}V_{ts}^*$.}

Examples of MFV models include models of supersymmetry with gauge- or
anomaly-mediation of its breaking.  If the LHC discovers new particles
that couple to the SM fermions, then it will be able to test solutions
to the new physics flavour puzzle such as MFV
\cite{Grossman:2007bd}. Much of its power to test such frameworks is
based on identifying top and bottom quarks.

To understand this statement, we note that the spurions $Y_U$ and
$Y_D$ can always be written in terms of the two diagonal Yukawa
matrices $\lambda_u$ and $\lambda_d$ and the CKM matrix $V$, see
\Erefs{speint} and (\ref{deflamd}). Thus, the only source of quark
flavour-changing transitions in MFV models is the CKM matrix. Next,
note that to an accuracy that is better than $\mathcal{O}(0.05)$, we
can write the CKM matrix as follows:
\begin{equation}\label{ckmapp}
V=\begin{pmatrix} 1&0.23&0\\ -0.23&1&0\\ 0&0&1\end{pmatrix}\SPp.
\end{equation}

\textbf{Exercise 9:} \emph{The approximation (\ref{ckmapp}) should be
intuitively obvious to top-physicists, but definitely
counter-intuitive to bottom-physicists. (Some of them have dedicated a
large part of their careers to experimental or theoretical efforts to
determine $V_{cb}$ and $V_{ub}$.) What does the approximation
imply for the bottom quark? When we take into account that it is
only good to $\mathcal{O}(0.05)$, what would the implications be?}

We learn that the third generation of quarks is decoupled, to a good
approximation, from the first two. This, in turn, means that any new
particle that couples to the SM quarks (think, for example, of heavy
quarks in vector-like representations of $G_\text{SM}$), decays into 
either a third-generation quark, or into a non-third-generation quark,
but not to both. For example, in \Bref{Grossman:2007bd}, MFV models
with additional charge $-1/3$, $SU(2)_\text{L}$-singlet quarks,
$B^\prime$, were considered. A concrete test of MFV was proposed,
based on the fact that the largest mixing effect involving the third
generation is of order $|V_{cb}|^2\sim0.002$: Is the following
prediction, concerning events of $B^\prime$ pair production,
fulfilled?
\begin{equation}
\frac{\Gamma(B^\prime\overline{B^\prime}\to Xq_{1,2}q_3)}
{\Gamma(B^\prime\overline{B^\prime}\to Xq_{1,2}q_{1,2})+
  \Gamma(B^\prime\overline{B^\prime}\to Xq_3q_3)}\lesssim10^{-3}.
\end{equation}
If not, then MFV is excluded.

\subsection{Supersymmetric flavour at the LHC}
One can think of analogous tests in the supersymmetric framework
\cite{Feng:2007ke,Engelhard:2009br, Feng:2009bs, Feng:2009yq, Hiller:2008wp,Hiller:2008sv}. Here, there is also a
generic prediction that, in each of the three sectors ($Q_L,U_R,D_R$),
squarks of
the first two generations are quasi-degenerate, and do not decay into
third-generation quarks. Squarks of the third generation can be
separated in mass (though, for small $\tan\beta$, the degeneracy in
the $\tilde D_R$ sector is threefold), and decay only to 
third-generation quarks.

It is not necessary, however, that the mediation of supersymmetry
breaking be MFV. Examples of natural and viable solutions to the
supersymmetric flavour problem that are not MFV include the following:
\begin{enumerate}
\item The leading contribution to the soft supersymmetry breaking
      terms is gauge mediated, and therefore MFV, but there are
      subleading contributions that are gravity mediated and provide
      new sources of flavour and CP violation
      \cite{Feng:2007ke,Hiller:2008sv}. The gravity mediated
      contributions could either have some structure (dictated, for
      example, by a Froggatt--Nielsen symmetry \cite{Feng:2007ke} or by
      localization in extra dimensions \cite{Nomura:2007ap}) or be
      anarchical \cite{HiHo2009}.
\item The first two sfermion generations are heavy, and their
      mixing with the third generation is suppressed (for a recent
      analysis, see Ref.~\cite{Giudice:2008uk}). These features can come,
      for example, from conformal dynamics \cite{Nelson:2000sn}.
\end{enumerate}

Such frameworks have different predictions concerning the mass
splitting between sfermion generations and the flavour decomposition
of the sfermion mass eigenstates. Note that measurements of 
flavour-changing neutral current processes are only sensitive to the products
of the form
\begin{equation}\label{eq:defdel}
\delta_{ij}=\frac{\Delta \tilde m^2_{ij}}{\tilde m^2}\ K_{ij}K_{jj}^*,
\end{equation}
where $\Delta\tilde m^2_{ij}$ is the mass-squared splitting between
the sfermion generations $i$ and $j$, $\tilde m^2$ is their average
mass-squared, and $K$ is the mixing matrix of gaugino couplings to
these sfermions. On the other hand, the LHC experiments---ATLAS and
CMS---can, at least in principle, measure the mass splitting and the
mixing separately~\cite{Feng:2009yq}.

\begin{figure}[htb]
\centering
\raisebox{70mm}{(a)}%
\includegraphics[width=.45\linewidth,clip]{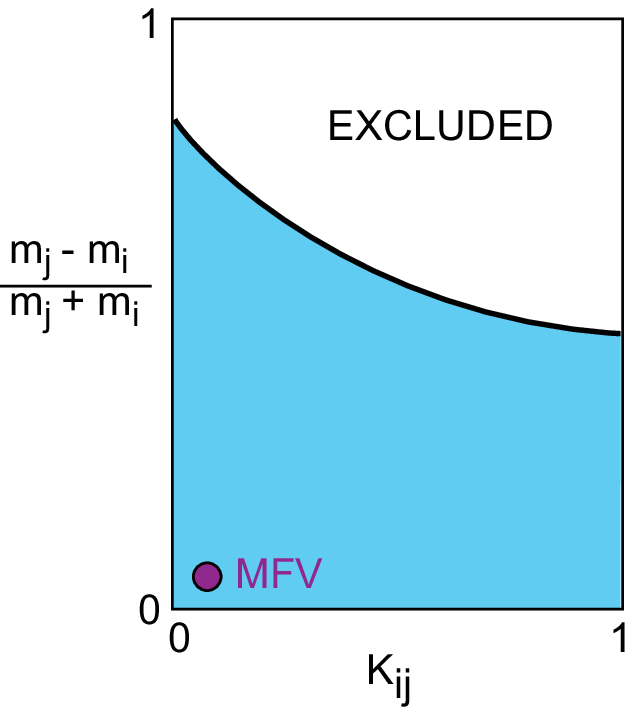}
\quad
\raisebox{70mm}{(b)}%
\includegraphics[width=.45\linewidth,clip]{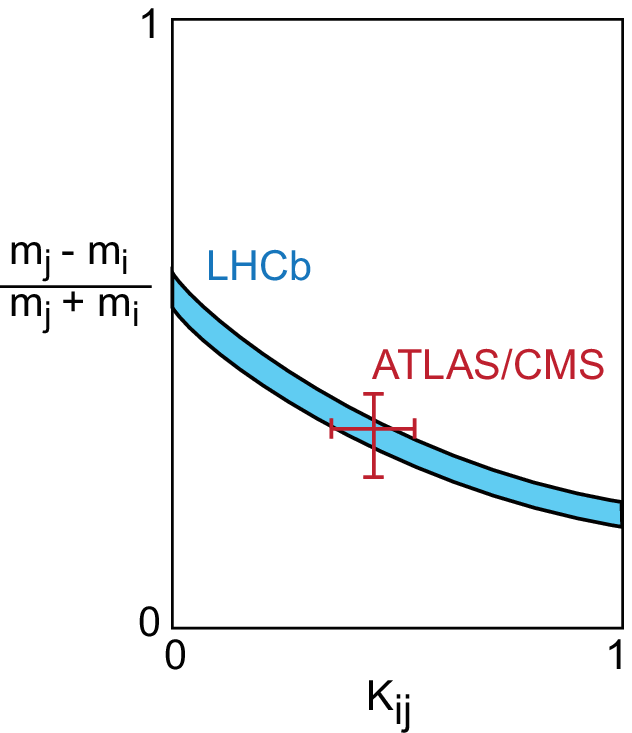}
\caption[]{Schematic description of the constraints in the plane of
           sfermion mass-squared splitting, $\Delta\tilde
           m^2_{ij}/\tilde m^2$, and mixing, $K_{ij}K_{jj}^*$: (a)
           Upper bounds from not observing any deviation from the SM
           predictions in present experiments; (b) Hypothetical future
           situation, where deviations have been observed in flavour
           factories (such as LHCb, a super-B factory, a $\mu\to
           e\gamma$ measurement, etc.) and the mass splitting and
           flavour decomposition have been measured by ATLAS/CMS.}
\label{fig:dmk}
\end{figure}

The present situation is depicted schematically in
\Fref{fig:dmk}(a). Flavour factories have provided only upper bounds
on deviations of FCNC processes, such as $\mu\to e\gamma$ or
$D^0$--$\overline{D}^0$ mixing, from the Standard Model
predictions. In the supersymmetric framework, such bounds translate
into an upper bound on a $\delta_{ij}$ parameter of \Eref{eq:defdel},
corresponding to the blue region in the figure. The supersymmetric
flavour puzzle can be stated as the question of why the region in the
upper right corner---where the flavour parameters are of order
one---is excluded. MFV often puts us in the lower left corner of the
plot, far from the experimental constraints (this is particularly true
for $\delta_{12}$ parameters).

The optimal future situation is depicted schematically in
\Fref{fig:dmk}(b). Imagine that a flavour factory does provide
evidence for new physics, such as observation of $\Gamma(\mu\to
e\gamma)\neq0$ or CP violation in $D^0$--$\overline{D}^0$ mixing. This
will constrain the corresponding $\delta$ parameter, which is shown as
the blue region in the figure. If ATLAS/CMS measure the corresponding
sfermion mass splitting and/or mixing, we shall get a small allowed
region in this flavour plane.

If we have at our disposal three such consistent measurements (rate of FCNC
process, spectrum and splitting), then we shall understand the
mechanism by which supersymmetry has its flavour violation
suppressed. This will provide strong hints about the mechanism of
supersymmetry breaking mediation.

If the sfermions are quasi-degenerate, then the mixing is determined
by the small corrections to the unit mass-squared matrix. As mentioned
above, the structure of such corrections may be dictated by the same
symmetry or dynamics that gives the structure of the Yukawa
couplings. If that is the case, then the measurement of the flavour
decomposition might shed light on the Standard Model flavour puzzle.

We conclude that measurements at the LHC related to new particles that
couple to the SM fermions are likely to teach us much more about flavour
physics.

\section{Neutrino anarchy versus quark hierarchy}
\label{sec:nu}

A detailed presentation of the physics and the formalism of neutrino
flavour transitions is given in Appendix \ref{sec:nufl} for both
vacuum oscillations (\ref{sec:vac}) and the matter transitions
(\ref{sec:mat}). It follows \Bref{Gonzalez-Garcia:2002dz}.

\textbf{Exercise 10:} \emph{For atmospheric $\nu_\mu$'s with $E\sim1\UGeV$, the
flux coming from above has $P_{\mu\mu}(L\sim10\Ukm)\approx1$, while
the flux from below has $P_{\mu\mu}(L\sim10^4\Ukm)\approx0.5$. Assuming
that for the flux coming from below the oscillations are averaged
out, estimate $\Delta m^2$ and $\sin^22\theta$.}

\textbf{Exercise 11:} \emph{For solar $\nu_e$'s, the transition
between matter ($\beta_\text{MSW}>1$) and vacuum
($\beta_\text{MSW}<\cos2\theta$) flavour transitions occurs around
$E\sim2\UMeV$. The transition probability is measured to be roughly
$P_{ee}\sim0.30$ for $\beta_\text{MSW}>1$. Estimate $\Delta m^2$ and
$\theta$ and predict $P_{ee}$ for $\beta_\text{MSW}\ll1$.}

The derived ranges for the three mixing angles and two mass-squared
differences at $1\sigma$ are \cite{GonzalezGarcia:2007ib}
\begin{eqnarray}\label{nupara}
\Delta m^2_{21}&=&(7.9\pm0.3)\times10^{-5}\UeV^2,\ \ \
|\Delta m^2_{32}|=(2.6\pm0.2)\times10^{-3}\UeV^2,\nonumber\\
\sin^2\theta_{12}&=&0.31\pm0.02,\ \ \
\sin^2\theta_{23}=0.47\pm0.07,\ \ \
\sin^2\theta_{13}=0^{+0.008}_{-0.0}.
\end{eqnarray}
The $3\sigma$ range for the matrix elements of $U$ are the following
\cite{GonzalezGarcia:2007ib}:
\begin{equation}\label{uthsi}
|U|=\begin{pmatrix}
   0.79\to0.86&0.50\to0.61&0.00\to0.20\\
   0.25\to0.53&0.47\to0.73&0.56\to0.79\\
   0.21\to0.51&0.42\to0.69&0.61\to0.83
    \end{pmatrix}\SPp.
\end{equation}

\subsection{New physics}
The simplest and most straightforward lesson of the evidence for
neutrino masses is also the most striking one: there is new physics
beyond the Standard Model. This is the first experimental result that
is inconsistent with the SM.

Most likely, the new physics is related to the existence of
$G_\text{SM}$-singlet fermions at some high energy scale that induce,
at low energies, the effective terms of \Eref{Hnint} through the
seesaw mechanism. The existence of heavy singlet fermions is predicted
by many extensions of the SM, especially by GUTs [beyond $SU(5)$] and
left--right-symmetric theories. The seesaw mechanism could also be
driven by an $SU(2)_\text{L}$-triplet fermion.

There are other possibilities. In particular, neutrino masses can be
generated without introducing any new fermions beyond those of the SM.
Instead, the existence of a scalar $\Delta_L(1,3)_{+1}$, that is, an
$SU(2)_\text{L}$-triplet, is required. The smallness of the neutrino
masses is related here to the smallness of the vacuum expectation
value $\langle\Delta_L^0\rangle$ (required also by the success of the
$\rho=1$ relation) and does not have a generic natural explanation.

In left--right-symmetric models, however, where the breaking of
$SU(2)_\text{R}\times U(1)_\text{B-L}\to U(1)_\text{Y}$ is induced by
the VEV of an $SU(2)_\text{R}$-triplet, $\Delta_R$, there must exist
also an $SU(2)_\text{L}$-triplet scalar. Furthermore, the Higgs
potential leads to an order of magnitude relation between the various
VEVs, $\langle\Delta_L^0\rangle\langle\Delta_R^0\rangle\sim v^2$, and
the smallness of $\langle\Delta_L^0\rangle$ is correlated with the
high scale of $SU(2)_\text{R}$ breaking. This situation can be thought
of as a seesaw of VEVs. In this model there are, however, also
SM-singlet fermions. The light neutrino masses arise from both the
seesaw mechanism (`type I') and the triplet VEV (`type II').

Neutrino masses could also be of the Dirac type. Here, again, singlet
fermions are introduced, but lepton number is imposed by hand. This
possibility is disfavoured by theorists since it is likely that global
symmetries are violated by gravitational effects. Furthermore, the
lightness of the neutrinos (compared to charged fermions) is
unexplained.

Another possibility is that neutrino masses are generated by mixing
with singlet fermions but the mass scale of these fermions is not
high. Here again the lightness of neutrino masses remains a
puzzle. The best known example of such a scenario is the framework of
supersymmetry without $R$ parity.

Let us emphasize that the seesaw mechanism or, more generally, the
extension of the SM with non-renormalizable terms, is the simplest
explanation of neutrino masses. Models in which neutrino masses are
generated by new physics at low energy imply a much more dramatic
departure from the SM. Furthermore, the existence of seesaw masses is
an unavoidable prediction of various extensions of the SM. In
contrast, many (but not all) of the low-energy mechanisms are
introduced for the specific purpose of generating neutrino masses.

\subsection{The scale of new physics}

\Eref[b]{Hnint} gives a light neutrino mass matrix:
\begin{equation}\label{seesawmass}
(M_\nu)_{ij}=Z_{ij}^\nu\frac{v^2}{\Lambda_\text{NP}}.
\end{equation}
It is straightforward to use the measured neutrino masses of
\Eref{nupara} in combination with \Eref{seesawmass} to estimate the
scale of new physics that is relevant to their generation. In
particular, if there is no quasi-degeneracy in the neutrino masses,
the heaviest of the active neutrino masses can be estimated:
\begin{equation}\label{mthree}
m_h=m_3\sim\sqrt{\Delta m^2_{32}}\approx0.05\UeV.
\end{equation}
(In the case of inverted hierarchy, the implied scale is
$m_h=m_2\sim\sqrt{\Delta m^2_{32}}\approx0.05\UeV$.) It follows that
the scale in the non-renormalizable terms (\ref{Hnint}) is given by
\begin{equation}\label{seesawlnp}
\Lambda_\text{NP}\sim v^2/m_h\approx10^{15}\UGeV.
\end{equation}
We should clarify two points regarding \Eref{seesawlnp}:
\begin{enumerate}
\item There could be some level of degeneracy between the neutrino
      masses. In such a case, \Eref{mthree} is modified into a lower
      bound on $m_3$ and, consequently, \Eref{seesawlnp} becomes an
      upper bound on $\Lambda_\text{NP}$.
\item It could be that the $Z_{ij}$ of \Eref{Hnint} are much smaller
      than 1. In such a case, again, \Eref{seesawlnp} becomes an upper
      bound on the scale of new physics.
\end{enumerate}

On the other hand, in models of approximate flavour symmetries, there
are relations between the structures of the charged lepton and
neutrino mass matrices that give, quite generically, $Z_{33}\gtrsim
m_\tau^2/v^2\sim10^{-4}$. We conclude that the likely range for
$\Lambda_\text{NP}$ is given by
\begin{equation}\label{lnpssfl}
10^{11}\UGeV\lesssim\Lambda_\text{NP}\lesssim10^{15}\UGeV\SPp.
\end{equation}

The estimates (\ref{seesawlnp}) and (\ref{lnpssfl}) are very
exciting. First, the upper bound on the scale of new physics is well
below the Planck scale. This means that there is new physics in Nature
which is intermediate between the two known scales, the Planck scale,
$m_\text{Pl}\sim10^{19}\UGeV$, and the electroweak breaking scale,
$v\sim 10^2\UGeV$.

Second, the scale $\Lambda_\text{NP}\sim10^{15}\UGeV$ is intriguingly
close to the scale of gauge coupling unification.

Third, the range (\ref{lnpssfl}) for the scale of lepton number
breaking is optimal for leptogenesis \cite{Fukugita:1986hr} (for
a recent review, see \Bref{Davidson:2008bu}). If
(i) leptogenesis is generated by the decays of the lightest singlet
neutrino $N_1$, and (ii) the masses of the singlet neutrinos are
hierarchical, $M_1/M_{2,3\ldots}\ll1$ , and (iii) the temperature
when leptogenesis occurs is high enough, $T_\text{LG}>10^{12}\UGeV$,
so that flavour effects are unimportant, then
there is an upper bound on the CP asymmetry in $N_1$ decays
\cite{Davidson:2002qv}:
\begin{equation}
|\epsilon_{N_1}|\leq\frac{3}{16\pi}\frac{M_1(m_3-m_2)}{v^2}.
\end{equation}
Given that $Y_B^\text{obs}\sim9\times10^{-11}$, and that
$Y_B\sim10^{-3}\eta\epsilon_{N_1}$, where $\eta\lesssim1$ is a washout
factor, we must require $|\epsilon_{N_1}|\gtrsim10^{-7}$. Moreover, we
have $m_3-m_2\leq\sqrt{\Delta m^2_{32}}\sim0.05\UeV$ and therefore
obtain $M_1\gtrsim10^{9}\UGeV$. Violating any of the three conditions
will relax this bound, but typically not by more than about an
order of magnitude.

\subsection{The flavour puzzle}
In the absence of neutrino masses, there are 13 flavour parameters in
the SM:
\begin{eqnarray}\label{chafla}
y_t&\sim&1,\ \ y_c\sim10^{-2},\ \ y_u\sim10^{-5},\nonumber\\
y_b&\sim&10^{-2},\ \ y_s\sim10^{-3},\ \ y_d\sim10^{-4},\nonumber\\
y_\tau&\sim&10^{-2},\ \ y_\mu\sim10^{-3},\ \ y_e\sim10^{-6},\nonumber\\
|V_{us}|&\sim&0.2,\ \ |V_{cb}|\sim0.04,\ \ |V_{ub}|\sim0.004,\ \
\sin\delta_\text{KM}\sim1.
\end{eqnarray}
These flavour parameters are hierarchical (their magnitudes span six
orders of magnitude), and all but two or three (the top Yukawa, the CP
violating phase, and perhaps the Cabibbo angle) are small. The
unexplained smallness and hierarchy pose the SM \emph{flavour puzzle}.
Its solution may direct us to physics beyond the Standard Model.

Several mechanisms have been proposed in response to this puzzle. For
example, approximate horizontal symmetries, broken by a small
parameter, can lead to selection rules that explain the hierarchy of
the Yukawa couplings.

In the extension of the SM with three active neutrinos that have
Majorana masses, there are nine new flavour parameters in addition to
those of \Eref{chafla}. These are three neutrino masses, three lepton
mixing angles, and three phases in the mixing matrix. Of the nine new
parameters, four have been measured: two mass-squared differences and
two mixing angles [see \Eref{nupara}]. This adds significantly to the
input data on flavour physics and provides an opportunity to test and
refine flavour models.

If neutrino masses arise from effective terms of the form of
\Eref{Hnint}, then the overall scale of neutrino masses is
related to the scale $\Lambda_\text{NP}$ and, in most cases, does not
tell us anything about flavour physics. More significant information
for flavour models can be written in terms of three dimensionless
parameters whose values can be read from \Eref{nupara}, that is
$\sin\theta_{12}$, $\sin\theta_{23}$ and
\begin{equation}\label{nuflpa}
\Delta m^2_{21}/|\Delta m^2_{32}|=0.030\pm0.003.
\end{equation}
In addition, the upper bound on $\sin\theta_{13}$ often plays a
significant role in flavour model building.

There are several features in the numerical estimates (\ref{nupara})
and (\ref{nuflpa}) that have drawn much attention and have driven
numerous investigations:

(i) \emph{Large mixing and strong hierarchy:} The mixing angle that is
relevant to the $2$--$3$ sector is large, $\sin\theta_{23}\sim0.7$. On
the other hand, if there is no quasi-degeneracy in the neutrino
masses, the corresponding mass ratio is small, $m_2/m_3\sim0.17$. It
is difficult to explain in a natural way a situation where there is an
$\mathcal{O}(1)$ mixing but the corresponding masses are hierarchical.

(ii) \emph{Two large and one small mixing angles:} The mixing angles relevant
to the $2$--$3$ sector ($\sin\theta_{23}\sim0.7$) and $1$--$2$ sector
($\sin\theta_{12}\sim0.55$) are large, yet the $1$--$3$ mixing angle is
small ($\sin\theta_{13}\lesssim 0.20$). Such a situation is, again,
difficult---though not impossible---to explain from approximate
symmetries. An example of a symmetry that does predict such a pattern
is that of $L_e$--$L_\mu$--$L_\tau$. This symmetry predicts, however,
$\theta_{12}\simeq\pi/4$, which is experimentally excluded.

(iii) \emph{Maximal mixing:} The value of $\theta_{23}$ is
intriguingly close to maximal mixing ($\sin^22\theta_{23}=1$). It is
interesting to understand whether a symmetry could explain this
special value.

(iv) \emph{Tribimaximal mixing:} The mixing matrix (\ref{uthsi}) has a
structure that is consistent with the following unitary matrix
\cite{Harrison:2002er}:
\begin{equation}
U=\begin{pmatrix}
   \sqrt{\frac23}&\sqrt{\frac13}&0\\
  -\sqrt{\frac16}&\sqrt{\frac13}&\sqrt{\frac{1}{2}}\\
   \sqrt{\frac16}&-\sqrt{\frac13}&\sqrt{\frac{1}{2}}
  \end{pmatrix}\SPp.
\end{equation}
It is interesting to understand whether a symmetry could explain this
special structure.

All four features enumerated above are difficult to explain in a large
class of flavour models that do very well in explaining the flavour
features of the quark sector. In particular, models with Abelian
horizontal symmetries (Froggatt--Nielsen type \cite{Froggatt:1978nt})
predict that, in general, $|V_{ub}|\sim|V_{us}V_{cb}|$, $|V_{ij}|\gtrsim
m_i/m_j$ ($i<j$) and $V\sim\mathbf{1}$
\cite{Leurer:1992wg,Leurer:1993gy}. All of these are successful
predictions. At the same time, however, these models predict
\cite{Grossman:1995hk} that for the neutrinos, in general,
$|U_{ij}|^2\sim m_i/m_j$ and $|U_{e3}|\sim|U_{e2}U_{\mu3}|$, in
contradiction to, respectively, points (i) and (ii) above (and there
is no way to make $\theta_{23}$ parametrically close to $\pi/4$). On
the other hand, there exist very specific models where these features
are related to a symmetry.

It is possible, however, that the above interpretation of the results
is wrong. Indeed, the data can be interpreted in a very different
way:

(v) \emph{No small parameters:} The two measured mixing angles are
larger than any of the quark mixing angles. Indeed, they are both of
order one. The measured mass ratio, $m_2/m_3\gtrsim0.16$ is larger
than any of the quark and charged lepton mass ratios, and could be
interpreted as an $\mathcal{O}(1)$ parameter (namely, it is
accidentally small, without any parametric suppression). If this is
the correct way of reading the data, the measured neutrino parameters
may actually reflect the absence of any hierarchical structure in the
neutrino mass matrices \cite{Hall:1999sn}. The possibility that there
is no structure---neither hierarchy, nor degeneracy---in the neutrino
sector has been called `neutrino mass anarchy'. An important test of
this idea will be provided by the measurement of $|U_{e3}|$. If indeed
the entries in $M_\nu$ have random values of the same order, all three
mixing angles are expected to be of order one. If experiments measure
$|U_{e3}|\sim0.1$, that is, close to the present bound, it can be
argued that its smallness is accidental. The stronger the upper bound
on this angle becomes, the more difficult it will be to maintain this
view.

Neutrino mass anarchy can be accommodated within models of Abelian
flavour symmetries, if the three lepton doublets carry the same
charge. Indeed, consider a supersymmetric model with a $U(1)_H$
symmetry that is broken by a single small spurion $\epsilon_H$ of
charge $-1$. Let us assume that the three fermion generations
contained in the $10$-representation of $SU(5)$ carry charges
$(2,1,0)$, while the three $\bar5$-representations carry charges
$(0,0,0)$. (The Higgs fields carry no $H$ charges.) Such a model
predicts $\epsilon_H^2$ hierarchy in the up sector, $\epsilon_H$
hierarchy in the down and charged lepton sectors, and anarchy in the
neutrino sector.

\textbf{Exercise 12:} \emph{The selection rule for this model is that
a term in the superpotential that carries $H$ charge $n\geq0$ is
suppressed by $\epsilon_H^n$. Find the parametric suppression of the
various entries in $M_u,M_d,M_\ell$, and $M_\nu$. Find the parametric
suppression of the mixing angles.}

It would be nice if the features of quark mass hierarchy and neutrino
mass anarchy can be traced back to some fundamental principle or to a
stringy origin (see, for example, \Bref{Antebi:2005hr}).

\section{Conclusions}
\label{sec:con}

\begin{enumerate}
\item[(i)] Measurements of CP violating $B$-meson decays have
           established that the Kobayashi--Maskawa mechanism is the
           dominant source of the observed CP violation.
\item[(ii)] Measurements of flavour-changing $B$-meson decays have
            established that the Cabibbo--Kobayashi--Maskawa mechanism
            is a major player in flavour violation.

\item[(iii)] The consistency of all these measurements with the CKM
             predictions sharpens the new physics flavour puzzle: If
             there is new physics at, or below, the \UTeVZ{} scale,
             then its flavour structure must be highly non-generic.

\item[(iv)] Measurements of $D^0$--$\overline{D}^0$ mixing imply that
            alignment by itself cannot solve the supersymmetric
            flavour problem.  The first two squark generations must be
            quasi-degenerate.

\item[(v)] Measurements of neutrino flavour parameters have not only
           not clarified the Standard Model flavour puzzle, but
           actually deepened it. Whether they imply an anarchical
           structure, or a tribimaximal mixing, it seems that the
           neutrino flavour structure is very different from that of
           quarks.

\item[(vi)] If the LHC experiments, ATLAS and CMS, discover new
            particles that couple to the Standard Model fermions,
            then, in principle, they will be able to measure new
            flavour parameters. Consequently, the new physics flavour
            puzzle is likely to be understood.

\item[(vii)] If the flavour structure of such new particles is
             affected by the same physics that sets the flavour
             structure of the Yukawa couplings, then the LHC
             experiments (and future flavour factories) may be able to
             shed light also on the Standard Model flavour puzzle.
\end{enumerate}
The huge progress in flavour physics in recent years has provided
answers to many questions. At the same time, new questions arise. We
look forward to the LHC era for more answers and more questions.

\section*{Acknowledgements}

The research of Y.~Nir is supported by the Israel Science Foundation;
the United States--Israel Binational Science Foundation (BSF),
Jerusalem, Israel; the German--Israeli Foundation for Scientific
Research and Development (GIF); and the Minerva Foundation.

\appendix
\section{The CKM matrix}
\label{app:ckm}
The CKM matrix $V$ is a $3\times3$ unitary matrix. Its form, however,
is not unique:

$(i)$ There is freedom in defining $V$ in that we can permute between
the various generations. This freedom is fixed by ordering the up quarks and
the down quarks by their masses, \ie $(u_1,u_2,u_3)\to(u,c,t)$ and
$(d_1,d_2,d_3)\to(d,s,b)$. The elements of $V$ are written as follows:
\begin{equation}\label{defVij}
V=\begin{pmatrix}
  V_{ud}&V_{us}&V_{ub}\\
  V_{cd}&V_{cs}&V_{cb}\\
  V_{td}&V_{ts}&V_{tb}
  \end{pmatrix}\SPp.
\end{equation}

$(ii)$ There is further freedom in the phase structure of $V$. This
means that the number of physical parameters in $V$ is smaller than
the number of parameters in a general unitary $3\times3$ matrix which
is nine (three real angles and six phases). Let us define $P_q$
($q=u,d$) to be diagonal unitary (phase) matrices. Then, if instead of
using $V_{qL}$ and $V_{qR}$ for the rotation (\ref{diagMq}) to the
mass basis we use $\tilde V_{qL}$ and $\tilde V_{qR}$, defined by
$\tilde V_{qL}=P_q V_{qL}$ and $\tilde V_{qR}=P_q V_{qR}$, we still
maintain a legitimate mass basis since $M_q^\text{diag}$ remains
unchanged by such transformations. However, $V$ does change:
\begin{equation}\label{eqphase}
V\to P_u V P_d^*.
\end{equation}
This freedom is fixed by demanding that $V$ has the minimal number of
phases. In the three-generation case $V$ has a single phase. (There
are five phase differences between the elements of $P_u$ and $P_d$ and,
therefore, five of the six phases in the CKM matrix can be removed.) This is
the Kobayashi--Maskawa phase $\delta_\text{KM}$ which is the single source of
CP violation in the quark sector of the Standard Model \cite{Kobayashi:1973fv}.

The fact that $V$ is unitary and depends on only four independent
physical parameters can be made manifest by choosing a specific
parametrization. The standard choice is \cite{Chau:1984fp}
\begin{equation}\label{stapar}
V=\begin{pmatrix}
  c_{12}c_{13}&s_{12}c_{13}&s_{13}e^{-i\delta}\\
 -s_{12}c_{23}-c_{12}s_{23}s_{13}e^{i\delta}&
  c_{12}c_{23}-s_{12}s_{23}s_{13}e^{i\delta}&s_{23}c_{13}\\
  s_{12}s_{23}-c_{12}c_{23}s_{13}e^{i\delta}&
 -c_{12}s_{23}-s_{12}c_{23}s_{13}e^{i\delta}&c_{23}c_{13}
  \end{pmatrix}\SPp,
\end{equation}
where $c_{ij}\equiv\cos\theta_{ij}$ and $s_{ij}\equiv\sin\theta_{ij}$.
The $\theta_{ij}$'s are the three real mixing parameters while
$\delta$ is the Kobayashi--Maskawa phase. It is known experimentally
that $s_{13}\ll s_{23}\ll s_{12}\ll1$. It is convenient to choose an
approximate expression where this hierarchy is manifest. This is the
Wolfenstein parametrization, where the four mixing parameters are
$(\lambda,A,\rho,\eta)$ with $\lambda=|V_{us}|=0.23$ playing the role
of an expansion parameter and $\eta$ representing the CP violating
phase \cite{Wolfenstein:1983yz,Buras:1994ec}:
\begin{equation}\label{wolpar}
V=\begin{pmatrix}
  1-\frac{1}{2}\lambda^2-\frac18\lambda^4 & \lambda & A\lambda^3(\rho-i\eta)\\
  -\lambda +\frac{1}{2}A^2\lambda^5[1-2(\rho+i\eta)] &
               1-\frac{1}{2}\lambda^2-\frac18\lambda^4(1+4A^2) & A\lambda^2 \\
  A\lambda^3[1-(1-\frac{1}{2}\lambda^2)(\rho+i\eta)]&
 -A\lambda^2+\frac{1}{2}A\lambda^4[1-2(\rho+i\eta)]& 1-\frac{1}{2}A^2\lambda^4
  \end{pmatrix}\SPp.
\end{equation}

\begin{figure}[tb]
  \centering
  \includegraphics[width=0.65\linewidth]{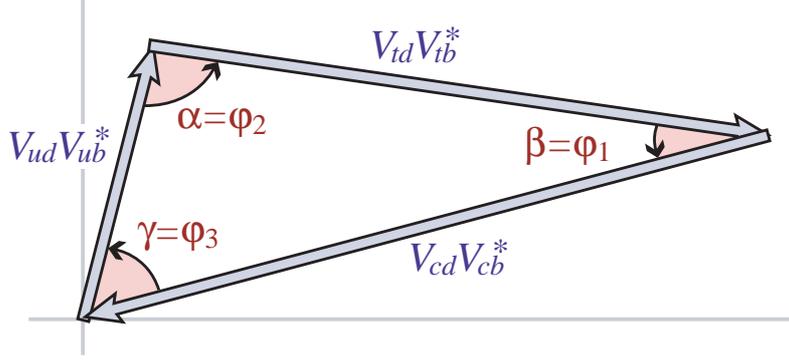}
  \caption[]{Graphical representation of the unitarity constraint
             $V_{ud}V_{ub}^*+V_{cd}V_{cb}^*+V_{td}V_{tb}^*=0$ as a
             triangle in the complex plane}
  \label{fg:tri}
\end{figure}

A very useful concept is that of the \emph{unitarity triangles}. The
unitarity of the CKM matrix leads to various relations among the
matrix elements, \eg
\begin{eqnarray}\label{Unitds}
V_{ud}V_{us}^*+V_{cd}V_{cs}^*+V_{td}V_{ts}^*=0,\\
\label{Unitsb}
V_{us}V_{ub}^*+V_{cs}V_{cb}^*+V_{ts}V_{tb}^*=0,\\
\label{Unitdb}
V_{ud}V_{ub}^*+V_{cd}V_{cb}^*+V_{td}V_{tb}^*=0.
\end{eqnarray}
Each of these three relations requires the sum of three complex
quantities to vanish and so can be geometrically represented in the
complex plane as a triangle. These are `the unitarity triangles',
though the term `unitarity triangle' is usually reserved for the
relation (\ref{Unitdb}) only. The unitarity triangle related to
\Eref{Unitdb} is depicted in \Fref{fg:tri}.

The rescaled unitarity triangle  is derived from (\ref{Unitdb})
by (a) choosing a phase convention such that $(V_{cd}V_{cb}^*)$
is real, and (b) dividing the lengths of all sides by $|V_{cd}V_{cb}^*|$.
Step (a) aligns one side of the triangle with the real axis, and
step (b) makes the length of this side 1. The form of the triangle
is unchanged. Two vertices of the rescaled unitarity triangle are
thus fixed at (0,0) and (1,0). The coordinates of the remaining
vertex correspond to the Wolfenstein parameters $(\rho,\eta)$.
The area of the rescaled unitarity triangle is $|\eta|/2$.

Depicting the rescaled unitarity triangle in the
$(\rho,\eta)$ plane, the lengths of the two complex sides are
\begin{equation}\label{RbRt}
R_u\equiv\left|\frac{V_{ud}V_{ub}}{V_{cd}V_{cb}}\right|
=\sqrt{\rho^2+\eta^2},\ \ \
R_t\equiv\left|\frac{V_{td}V_{tb}}{V_{cd}V_{cb}}\right|
=\sqrt{(1-\rho)^2+\eta^2}.
\end{equation}
The three angles of the unitarity triangle are defined as follows
\cite{Dib:1989uz,Rosner:1988nx}:
\begin{equation}\label{abcangles}
\alpha\equiv\arg\left[-\frac{V_{td}V_{tb}^*}{V_{ud}V_{ub}^*}\right],\quad
\beta \equiv\arg\left[-\frac{V_{cd}V_{cb}^*}{V_{td}V_{tb}^*}\right],\quad
\gamma\equiv\arg\left[-\frac{V_{ud}V_{ub}^*}{V_{cd}V_{cb}^*}\right]\SPp.
\end{equation}
They are physical quantities and can be independently measured by CP
asymmetries in $B$ decays. It is also useful to define the two
small angles of the unitarity triangles (\ref{Unitds}), (\ref{Unitsb}):
\begin{equation}\label{bbangles}
\beta_s\equiv\arg\left[-\frac{V_{ts}V_{tb}^*}{V_{cs}V_{cb}^*}\right],\quad
\beta_K\equiv\arg\left[-\frac{V_{cs}V_{cd}^*}{V_{us}V_{ud}^*}\right]\SPp.
\end{equation}

The $\lambda$ and $A$ parameters are very well determined at present,
see \Eref{lamaexp}. The main effort in CKM measurements is thus
aimed at improving our knowledge of $\rho$ and $\eta$:
\begin{equation}
\rho=0.14^{+0.03}_{-0.02},\ \ \ \eta=0.35\pm0.02.
\end{equation}
The present status of our knowledge is best seen in a plot of the
various constraints and the final allowed region in the $\rho$--$\eta$
plane. This is shown in \Fref{fg:UT}.

\section{CP violation in neutral $B$ decays to final CP eigenstates}
\label{sec:formalism}

We define decay amplitudes of $B$ (which could be charged or neutral)
and its CP conjugate $\Bbar$ to a multiparticle final state $f$ and its
CP conjugate $\fb$ as
\begin{equation}\label{decamp}
A_{\f}=\langle \f|\mathcal{H}|B\rangle\quad , \quad
\overline{A}_{\f}=\langle \f|\mathcal{H}|\Bbar\rangle\quad , \quad
A_{\fb}=\langle \fb|\mathcal{H}|B\rangle\quad , \quad
\overline{A}_{\fb}=\langle \fb|\mathcal{H}|\Bbar\rangle\; ,
\end{equation}
where $\mathcal{H}$ is the Hamiltonian governing
weak interactions.  The action of CP on these states introduces
phases $\xi_B$ and $\xi_f$ according to
\begin{eqnarray}\label{eq:phaseconv}
\CP|B\rangle &=& e^{+i\xi_{B}}\,|\Bbar\rangle \quad , \quad
\CP|\f\rangle = e^{+i\xi_{\f}}\,|\fb\rangle \; ,\nonumber\\
\CP|\Bbar\rangle& =& e^{-i\xi_{B}}\,|B\rangle \quad , \quad
\CP|\fb\rangle = e^{-i\xi_{\f}}\,|\f\rangle \ ,
\end{eqnarray}
so that $(\CP)^2=1$. The phases $\xi_B$ and $\xi_f$ are arbitrary and
unphysical because of the flavour symmetry of the strong
interaction. If CP is conserved by the dynamics, $[\CP,\mathcal{H}] =
0$, then $A_f$ and $\overline{A}_{\fb}$ have the same magnitude and an
arbitrary unphysical relative phase
\begin{equation}\label{spupha}
\overline{A}_{\fb} = e^{i(\xi_{\f}-\xi_{B})}\, A_f\; .
\end{equation}

A state that is initially a superposition of $\Bz$ and $\Bzb$, say
\begin{equation}
|\psi(0)\rangle = a(0)|\Bz\rangle+b(0)|\Bzb\rangle \; ,
\end{equation}
will evolve in time acquiring components that describe all possible
decay final states $\{f_1,f_2,\ldots\}$, that is,
\begin{equation}
|\psi(t)\rangle =
a(t)|\Bz\rangle+b(t)|\Bzb\rangle+c_1(t)|f_1\rangle+c_2(t)|f_2\rangle+\cdots
\; .
\end{equation}
If we are interested in computing only the values of $a(t)$ and $b(t)$
(and not the values of all $c_i(t)$), and if the times $t$ in which we
are interested are much larger than the typical strong interaction
scale, then we can use a much simplified
formalism~\cite{Weisskopf:au}. The simplified time evolution is
determined by a $2\times 2$ effective Hamiltonian $\Heff$ that is
not Hermitian, since otherwise the mesons would only oscillate and not
decay. Any complex matrix, such as $\Heff$, can be written in terms of
Hermitian matrices $\Meff$ and $\Geff$ as
\begin{equation}
\Heff = \Meff - \frac{i}{2}\,\Geff \; .
\end{equation}
$\Meff$ and $\Geff$ are associated with
$(\Bz,\Bzb)\leftrightarrow(\Bz,\Bzb)$ transitions via off-shell
(dispersive) and on-shell (absorptive) intermediate states, respectively.
Diagonal elements of $\Meff$ and $\Geff$ are associated with the
flavour-conserving transitions $\Bz\to\Bz$ and $\Bzb\to\Bzb$ while
off-diagonal elements are associated with flavour-changing transitions
$\Bz\leftrightarrow\Bzb$.

The eigenvectors of $\Heff$ have well-defined masses and decay
widths. We introduce complex parameters $p_{L,H}$ and $q_{L,H}$ to
specify the components of the strong interaction eigenstates, $\Bz$ and
$\Bzb$, in the light ($B_L$) and heavy ($B_H$) mass eigenstates:
\begin{equation}\label{defpq}
|B_{L,H}\rangle=p_{L,H}|\Bz\rangle\pm q_{L,H}|\Bzb\rangle
\end{equation}
with the normalization $|p_{L,H}|^2+|q_{L,H}|^2=1$. If either CP or
CPT is a symmetry of $\Heff$ (independently of whether T is conserved or
violated) then $\Meff_{11} = \Meff_{22}$ and $\Geff_{11}=
\Geff_{22}$, and solving the eigenvalue problem for $\Heff$ yields $p_L
= p_H \equiv p$ and $q_L = q_H \equiv q$ with
\begin{equation}
\left(\frac{q}{p}\right)^2=\frac{\Meff_{12}^\ast -
    (i/2)\Geff_{12}^\ast}{\Meff_{12}-(i/2)\Geff_{12}}\; .
\end{equation}
From now on we assume that CPT is conserved.
If either CP or T is a symmetry of $\Heff$ (independently of whether
CPT is conserved or violated), then $\Meff_{12}$ and $\Geff_{12}$ are
relatively real, leading to
\begin{equation}
\left(\frac{q}{p}\right)^2 = e^{2i\xi_B} \quad \Rightarrow \quad
\left|\frac{q}{p}\right| = 1 \; ,
\end{equation}
where $\xi_B$ is the arbitrary unphysical phase introduced in
\Eref{eq:phaseconv}.

The real and imaginary parts of the eigenvalues of $\Heff$
corresponding to $|B_{L,H}\rangle$ represent their masses and
decay-widths, respectively. The mass difference $\Delta m_B$ and the
width difference $\Delta\Gamma_B$ are defined as follows:
\begin{equation}\label{DelmG}
\Delta m_B\equiv M_H-M_L,\quad\Delta\Gamma_B\equiv\Gamma_H-\Gamma_L\SPp.
\end{equation}
Note that here $\Delta m_B$ is positive by definition, while the sign of
$\Delta\Gamma_B$ is to be experimentally determined.
The average mass and width are given by
\begin{equation}\label{aveMG}
m_B\equiv\frac{M_H+M_L}{2},\quad\Gamma_B\equiv\frac{\Gamma_H+\Gamma_L}{2}\SPp.
\end{equation}
It is useful to define dimensionless ratios $x$ and $y$:
\begin{equation}\label{defxy}
x\equiv\frac{\Delta m_B}{\Gamma_B},\quad y\equiv\frac{\Delta\Gamma_B}{2\Gamma_B}.
\end{equation}
Solving the eigenvalue equation gives
\begin{equation}\label{eveq}
(\Delta m_B)^2-\frac{1}{4}(\Delta\Gamma_B)^2=(4|M_{12}|^2-|\Gamma_{12}|^2),\ \ \ \
\Delta m_B\Delta\Gamma_B=4\re{M_{12}\Gamma_{12}^*}.
\end{equation}

All CP-violating observables in $B$ and $\Bbar$ decays to final states $f$
and $\fb$ can be expressed in terms of phase-convention-independent
combinations of $A_f$, $\overline{A}_f$, $A_{\overline{f}}$, and
$\overline{A}_{\overline{f}}$, together with, for neutral-meson decays
only, $q/p$. CP violation in charged-meson decays depends only on the
combination $|\overline{A}_{\fb}/A_f|$, while CP violation in
neutral-meson decays is complicated by $\Bz\leftrightarrow\Bzb$
oscillations and depends, additionally, on $|q/p|$ and on $\lambda_f
\equiv (q/p)(\overline{A}_f/A_f)$.

For neutral $D$, $B$, and $B_s$ mesons, $\Delta\Gamma/\Gamma\ll1$ and
so both mass eigenstates must be considered in their evolution. We
denote the state of an initially pure $|\Bz\rangle$ or $|\Bzb\rangle$
after an elapsed proper time $t$ as $|\Bz_{\mathrm{phys}}(t)\rangle$
or $|\Bzb_{\mathrm{phys}}(t)\rangle$, respectively. Using the
effective Hamiltonian approximation, we obtain
\begin{eqnarray}\label{defphys}
|\Bz_\text{phys}(t)\rangle&=&g_+(t)\,|\Bz\rangle
- \frac qp\ g_-(t)|\Bzb\rangle,\nonumber\\
|\Bzb_\text{phys}(t)\rangle&=&g_+(t)\,|\Bzb\rangle
- \frac pq\ g_-(t)|\Bz\rangle \; ,
\end{eqnarray}
where
\begin{equation}
g_\pm(t) \equiv \frac{1}{2}\left(e^{-im_Ht-\frac{1}{2}\Gamma_Ht}\pm
  e^{-im_Lt-\frac{1}{2}\Gamma_Lt}\right).
\end{equation}

One obtains the following time-dependent decay rates:
\begin{eqnarray}
\frac{d\Gamma[\Bz_\text{phys}(t)\to f]/dt}{e^{-\Gamma t}\mathcal{N}_f}&=&
\left(|A_f|^2+|(q/p)\overline{A}_f|^2\right)\cosh(y\Gamma t)
  +\left(|A_f|^2-|(q/p)\overline{A}_f|^2\right)\cos(x\Gamma t)\nonumber\\
&+&2\,\re{(q/p)A_f^\ast \overline{A}_f}\sinh(y\Gamma t)
-2\,\im{(q/p)A_f^\ast \overline{A}_f}\sin(x\Gamma t)
\label{decratbt1}\;,\\
\frac{d\Gamma[\Bzb_\text{phys}(t)\to f]/dt}{e^{-\Gamma t}\mathcal{N}_f}&=&
\left(|(p/q)A_f|^2+|\overline{A}_f|^2\right)\cosh(y\Gamma t)
  -\left(|(p/q)A_f|^2-|\overline{A}_f|^2\right)\cos(x\Gamma t)\nonumber\\
&+&2\,\re{(p/q)A_f\overline{A}^\ast_f}\sinh(y\Gamma t)
-2\,\im{(p/q)A_f\overline{A}^\ast_f}\sin(x\Gamma t)
\label{decratbt2}\; ,
\end{eqnarray}
where $\mathcal{N}_f$ is a common normalization factor. Decay rates to
the CP-conjugate final state $\fb$ are obtained analogously, with
$\mathcal{N}_f = \mathcal{N}_{\fb}$ and the substitutions $A_f\to
A_{\fb}$ and $\overline{A}_f\to\overline{A}_{\fb}$ in
\Erefs{decratbt1} and (\ref{decratbt2}). Terms proportional to
$|A_f|^2$ or $|\overline{A}_f|^2 $ are associated with decays that
occur without any net $B\leftrightarrow\Bbar$ oscillation, while terms
proportional to $|(q/p)\overline{A}_f|^2$ or $|(p/q)A_f|^2$ are
associated with decays following a net oscillation. The $\sinh(y\Gamma
t)$ and $\sin(x\Gamma t)$ terms of \Erefs{decratbt1} and
(\ref{decratbt2}) are associated with the interference between these
two cases. Note that, in multi-body decays, amplitudes are functions
of phase-space variables. Interference may be present in some regions
but not in others, and is strongly influenced by resonant substructure.

One possible manifestation of CP-violating effects in meson decays
\cite{Nir:1992uv} is in the interference between a decay without
mixing, $\Bz\to f$, and a decay with mixing, $\Bz\to \Bzb\to f$ (such
an effect occurs only in decays to final states that are common to
$\Bz$ and $\Bzb$, including all CP eigenstates). It is defined by
\begin{equation}\label{cpvint}
\im{\lambda_f}\ne 0 \; ,
\end{equation}
with
\begin{equation}\label{deflam}
\lambda_f \equiv \frac{q}{p}\frac{\overline{A}_f}{A_f} \; .
\end{equation}
This form of CP violation can be observed, for example, using the
asymmetry of neutral meson decays into final CP eigenstates $f_{\CP}$
\begin{equation}\label{asyfcp}
\mathcal{A}_{f_{\CP}}(t)\equiv\frac{d\Gamma/dt[\Bzb_\text{phys}(t)\to f_{\CP}]-
d\Gamma/dt[\Bz_\text{phys}(t)\to f_{\CP}]}
{d\Gamma/dt[\Bzb_\text{phys}(t)\to f_{\CP}]+d\Gamma/dt[\Bz_\text{phys}(t)\to
  f_{\CP}]}\; .
\end{equation}
For $\Delta\Gamma = 0$ and $|q/p|=1$ (which is a good approximation
for $B$ mesons), $\mathcal{A}_{f_{\CP}}$ has a particularly simple form
\cite{Dunietz:1986vi,Blinov:ru,Bigi:1986vr}:
\begin{eqnarray}\label{asyfcpb}
\mathcal{A}_{f}(t)&=&S_f\sin(\Delta mt)-C_f\cos(\Delta mt),\nonumber\\
S_f&\equiv&\frac{2\,\im{\lambda_{f}}}{1+|\lambda_{f}|^2},\ \ \
C_f\equiv\frac{1-|\lambda_{f}|^2}{1+|\lambda_{f}|^2}\SPp.
\end{eqnarray}

Consider the $B\to f$ decay amplitude $A_f$, and the CP conjugate
process $\Bbar\to\fb$ with decay amplitude $\overline{A}_{\fb}$. There
are two types of phases that may appear in these decay amplitudes.
Complex parameters in any Lagrangian term that contributes to the
amplitude will appear in complex conjugate form in the CP-conjugate
amplitude. Thus their phases appear in $A_f$ and
$\overline{A}_{\overline{f}}$ with opposite signs. In the Standard
Model, these phases occur only in the couplings of the $W^\pm$ bosons
and hence are often called `weak phases'. The weak phase of any
single term is convention dependent. However, the difference between
the weak phases in two different terms in $A_f$ is convention
independent. A second type of phase can appear in scattering or decay
amplitudes even when the Lagrangian is real. Their origin is the
possible contribution from intermediate on-shell states in the decay
process. Since these phases are generated by CP-invariant
interactions, they are the same in $A_f$ and
$\overline{A}_{\overline{f}}$. Usually the dominant rescattering is
due to strong interactions and hence the designation `strong phases'
for the phase shifts so induced. Again, only the relative strong
phases between different terms in the amplitude are physically
meaningful.

The `weak' and `strong' phases discussed here appear in addition to
the `spurious' CP~transformation phases of \Eref{spupha}. Those
spurious phases are due to an arbitrary choice of phase convention,
and do not originate from any dynamics or induce any \CP
violation. For simplicity, we set them to zero from here on.

It is useful to write each contribution $a_i$ to $A_f$ in three parts:
its magnitude $|a_i|$, its weak phase $\phi_i$, and its strong
phase $\delta_i$. If, for example, there are two such
contributions, $A_f = a_1 + a_2$, we have
\begin{eqnarray}\label{weastr}
A_f&=& |a_1|e^{i(\delta_1+\phi_1)}+|a_2|e^{i(\delta_2+\phi_2)},\nonumber\\
\overline{A}_{\overline{f}}&=&
|a_1|e^{i(\delta_1-\phi_1)}+|a_2|e^{i(\delta_2-\phi_2)}.
\end{eqnarray}
Similarly, for neutral meson decays, it is useful to write
\begin{equation}\label{defmgam}
\Meff_{12} = |\Meff_{12}| e^{i\phi_M} \quad , \quad
\Geff_{12} = |\Geff_{12}| e^{i\phi_\Gamma} \; .
\end{equation}
Each of the phases appearing in \Erefs{weastr} and (\ref{defmgam}) is
convention dependent, but combinations such as $\delta_1-\delta_2$,
$\phi_1-\phi_2$, $\phi_M-\phi_\Gamma$ and
$\phi_M+\phi_1-\overline{\phi}_1$ (where $\overline{\phi}_1$ is a weak
phase contributing to $\overline{A}_f$) are physical.

In the approximations that only a single weak phase contributes to decay,
$A_f=|a_f|e^{i(\delta_f+\phi_f)}$, and that
$|\Geff_{12}/\Meff_{12}|=0$, we obtain $|\lambda_f|=1$ and
the \CP asymmetries in decays to a final CP
eigenstate $f$ [\Eref{asyfcp}] with eigenvalue $\eta_f= \pm 1$
are given by
\begin{equation}\label{afcth}
\mathcal{A}_{f_{\CP}}(t) = \im{\lambda_f}\; \sin(\Delta m t) \; \
\mathrm{with}\ \
\im{\lambda_f}=\eta_f\sin(\phi_M+2\phi_f).
\end{equation}
Note that the phase so measured is purely a weak phase, and no
hadronic parameters are involved in the extraction of its value from
$\im{\lambda_f}$.

\section{Supersymmetric contributions to neutral meson mixing}
\label{app:susyd}

We consider the squark--gluino box diagram contribution to
$D^0$--$\overline{D}^0$ mixing amplitude that is proportional to
$K_{2i}^u K^{u*}_{1i}K_{2j}^u K^{u*}_{1j}$, where $K^u$ is the mixing
matrix of the gluino couplings to left-handed up quarks and their up
squark partners. (In the language of the mass insertion approximation,
we calculate here the contribution that is $\propto
[(\delta^u_{LL})_{12}]^2$.) We work in the mass basis for both quarks
and squarks.

The contribution is given by
\begin{equation}\label{motsusy}
M_{12}^D=-i\frac{4\pi^2}{27}\alpha_s^2m_Df_D^2B_D\eta_\text{QCD}
\sum_{i,j}(K_{2i}^uK_{1i}^{u*}K_{2j}^uK_{1j}^{u*})(11\tilde
I_{4ij}+4\tilde m_g^2I_{4ij})\SPp,
\end{equation}
where
\begin{eqnarray}
\tilde I_{4ij}&\equiv&\int\frac{d^4p}{(2\pi)^4}\frac{p^2}{(p^2-\tilde
  m_g^2)^2(p^2-\tilde m_i^2)(p^2-\tilde m_j^2)}\nonumber\\
&=&\frac{i}{(4\pi)^2}\left[\frac{\tilde m_g^2}
  {(\tilde m_i^2-\tilde m_g^2)(\tilde m_j^2-\tilde m_g^2)}\right.\nonumber\\
   && +\left.\frac{\tilde m_i^4}
  {(\tilde m_i^2-\tilde m_j^2)(\tilde m_i^2-\tilde
    m_g^2)^2}\ln\frac{\tilde m_i^2}{\tilde m_g^2}
  +\frac{\tilde m_j^4}
  {(\tilde m_j^2-\tilde m_i^2)(\tilde m_j^2-\tilde
    m_g^2)^2}\ln\frac{\tilde m_j^2}{\tilde m_g^2}\right],
\end{eqnarray}
\begin{eqnarray}
I_{4ij}&\equiv&\int\frac{d^4p}{(2\pi)^4}\frac{1}{(p^2-\tilde
  m_g^2)^2(p^2-\tilde m_i^2)(p^2-\tilde m_j^2)}\nonumber\\
&=&\frac{i}{(4\pi)^2}\left[\frac{1}
  {(\tilde m_i^2-\tilde m_g^2)(\tilde m_j^2-\tilde m_g^2)}\right.\nonumber\\
   && +\left.\frac{\tilde m_i^2}
  {(\tilde m_i^2-\tilde m_j^2)(\tilde m_i^2-\tilde
    m_g^2)^2}\ln\frac{\tilde m_i^2}{\tilde m_g^2}
  +\frac{\tilde m_j^2}
  {(\tilde m_j^2-\tilde m_i^2)(\tilde m_j^2-\tilde
    m_g^2)^2}\ln\frac{\tilde m_j^2}{\tilde m_g^2}\right].
\end{eqnarray}

We now follow the discussion in \Brefs{Raz:2002zx,Nir:2002ah}.
To see the consequences of the super-GIM mechanism, let us expand the
expression for the box integral around some value $\tilde m^2_q$ for
the squark masses-squared:
\begin{eqnarray}
I_4(\tilde m_g^2,\tilde m_i^2,\tilde m_j^2)&=&
I_4(\tilde m_g^2,\tilde m_q^2+\delta\tilde m_i^2,\tilde
m_q^2+\delta\tilde m_j^2)\nonumber\\
&=&I_4(\tilde m_g^2,\tilde m_q^2,\tilde m_q^2)
+(\delta\tilde m_i^2+\delta\tilde m_j^2)I_5(\tilde m_g^2,\tilde
m_q^2,\tilde m_q^2,\tilde m_q^2)\nonumber\\
&+&\frac{1}{2}\left[(\delta\tilde m_i^2)^2+(\delta\tilde
  m_j^2)^2+2(\delta\tilde m_i^2)(\delta\tilde m_j^2)\right]I_6(\tilde m_g^2,\tilde
m_q^2,\tilde m_q^2,\tilde m_q^2,\tilde m_q^2)+\cdots
\end{eqnarray}
where
\begin{equation}
I_n(\tilde m_g^2,\tilde m_q^2,\ldots,\tilde
m_q^2)\equiv\int\frac{d^4p}{(2\pi)^4}\frac{1}{(p^2-\tilde
  m_g^2)^2(p^2-\tilde m_q^2)^{n-2}},
\end{equation}
and similarly for $\tilde I_{4ij}$. Note that $I_n\propto(\tilde
m_q^2)^{n-2}$ and $\tilde I_n\propto(\tilde m_q^2)^{n-3}$. Thus, using
$x\equiv\tilde m_g^2/\tilde m_q^2$, it is customary to define
\begin{equation}
I_n\equiv\frac{i}{(4\pi)^2(\tilde m_q^2)^{n-2}}f_n(x),\ \ \ \
\tilde I_n\equiv\frac{i}{(4\pi)^2(\tilde m_q^2)^{n-3}}\tilde f_n(x).
\end{equation}
The unitarity of the mixing matrix implies that
\begin{equation}
\sum_i (K_{2i}^uK_{1i}^{u*}K_{2j}^uK_{1j}^{u*})=
\sum_j (K_{2i}^uK_{1i}^{u*}K_{2j}^uK_{1j}^{u*})=0.
\end{equation}
We learn that the terms that are proportional $f_4,\tilde f_4,f_5$, and
$\tilde f_5$ vanish in their contribution to $M_{12}$. When
$\delta\tilde m_i^2\ll\tilde m_q^2$ for all $i$, the
leading contributions to $M_{12}$ come from $f_6$ and $\tilde f_6$. We
learn that for quasi-degenerate squarks, the leading contribution is
quadratic in the small mass-squared difference. The functions $f_6(x)$
and $\tilde f_6(x)$ are given by
\begin{eqnarray}
f_6(x)&=&\frac{6(1+3x)\ln x+x^3-9x^2-9x+17}{6(1-x)^5},\nonumber\\
\tilde f_6(x)&=&\frac{6x(1+x)\ln x-x^3-9x^2+9x+1}{3(1-x)^5}.
\end{eqnarray}
For example, with $x=1$, $f_6(1)=-1/20$ and $\tilde f_6=+1/30$;
with $x=2.33$, $f_6(2.33)=-0.015$ and $\tilde f_6=+0.013$.

To further simplify things, let us consider a two-generation
case. Then
\begin{eqnarray}
M_{12}^D&\propto& 2(K_{21}^uK_{11}^{u*})^2(\delta\tilde
m_1^2)^2+2(K_{22}^uK_{12}^{u*})^2(\delta\tilde
m_2^2)^2+(K_{21}^uK_{11}^{u*}K_{22}^uK_{12}^{u*})(\delta\tilde
m_1^2+\delta\tilde m_2^2)^2\nonumber\\
&=&(K^u_{21}K_{11}^{u*})^2(\tilde m_2^2-\tilde m_1^2)^2.
\end{eqnarray}
We thus rewrite \Eref{motsusy} for the case of quasi-degenerate
squarks:
\begin{equation}\label{motsusyd}
M_{12}^D=\frac{\alpha_s^2m_Df_D^2B_D\eta_\text{QCD}}{108\tilde m_q^2}
[11\tilde f_6(x)+4xf_6(x)]\frac{(\Delta\tilde m^2_{21})^2}{\tilde m_q^4}
(K_{21}^uK_{11}^{u*})^2.
\end{equation}
For example, for $x=1$, $11\tilde f_6(x)+4xf_6(x)=+0.17$.
For $x=2.33$, $11\tilde f_6(x)+4xf_6(x)=+0.003$.

\section{Neutrino flavour transitions}
\label{sec:nufl}
\subsection{Neutrinos in vacuum}
\label{sec:vac}
Neutrino oscillations in vacuum \cite{Pontecorvo:1957cp} arise since
neutrinos are massive and mix. In other words, the neutrino state that
is produced by electroweak interactions is not a mass eigenstate.
The weak eigenstates $\nu_\alpha$ ($\alpha=e,\mu,\tau$ denotes the
charged lepton mass eigenstates and their neutrino doublet-partners)
are linear combinations of the mass eigenstates $\nu_i$ ($i=1,2,3$):
\begin{equation}
|\nu_\alpha\rangle=U_{\alpha i}^*|\nu_i\rangle.
\end{equation}
After travelling a distance $L$ (or, equivalently for relativistic
neutrinos, time $t$), a neutrino originally produced with a flavour
$\alpha$ evolves as follows:
\begin{equation}
|\nu_\alpha(t)\rangle=U_{\alpha i}^*|\nu_i(t)\rangle.
\end{equation}
It can be detected in the charged-current interaction
$\nu_\alpha(t)N^\prime\to\ell_\beta N$ with a probability
\begin{equation}
P_{\alpha\beta}=|\langle\nu_\beta|\nu_\alpha(t)\rangle|^2=
\left|\sum_{i=1}^3\sum_{j=1}^3U_{\alpha i}^*U_{\beta
    j}\langle\nu_j(0)|\nu_i(t)\rangle\right|^2.
\end{equation}
We follow the analysis of \Bref{Gonzalez-Garcia:2002dz}.  We use the
standard approximation that $|\nu\rangle$ is a plane wave,
$|\nu_i(t)\rangle=e^{-iE_it}|\nu_i(0)\rangle$. In all cases of
interest to us, the neutrinos are relativistic:
\begin{equation}
E_i=\sqrt{p_i^2+m_i^2}\simeq p_i+\frac{m_i^2}{2E_i},
\end{equation}
where $E_i$ and $m_i$ are, respectively, the energy and the mass of
the neutrino mass eigenstate. Furthermore, we can assume that
$p_i\simeq p_j\equiv p\simeq E$. Then, we obtain the following
transition probability:
\begin{equation}\label{palbe}
P_{\alpha\beta}=\delta_{\alpha\beta}-4\sum_{i=1}^2\sum_{j=i+1}^3\mathcal{R}e
    \left(U_{\alpha i}U_{\beta i}^*U_{\alpha j}^*U_{\beta
    j}\right)\sin^2 x_{ij},
\end{equation}
where $x_{ij}\equiv\Delta m^2_{ij}L/(4E)$, $\Delta
m^2_{ij}=m_i^2-m_j^2$, and $L=t$ is the distance between the source
(that is, the production point of $\nu_\alpha$) and the detector (that
is, the detection point of $\nu_\beta$). In deriving \Eref{palbe}
we used the orthogonality relation $\langle\nu_j(0)|\nu_i(0)\rangle
=\delta_{ij}$. It is convenient to use the following units:
\begin{equation}
x_{ij}=1.27\ \frac{\Delta m^2_{ij}}{\UeVZ^2}\ \frac{L/E}{\textrm{m}/\UMeVZ}.
\end{equation}
The transition probability [\Eref{palbe}] has an oscillatory
behaviour, with oscillation lengths
\begin{equation}
L_{0,ij}^\text{osc}=\frac{4\pi E}{\Delta m^2_{ij}}
\end{equation}
and amplitude that is proportional to elements of the mixing
matrix. Thus, in order to have oscillations, neutrinos must have
different masses ($\Delta m^2_{ij}\neq0$) and they must mix
($U_{\alpha i}U_{\beta i}\neq 0$).

An experiment is characterized by the typical neutrino energy $E$ and
by the source-detector distance $L$. In order to be sensitive to a
given value of $\Delta m^2_{ij}$, the experiment has to be set up with
$E/L\approx\Delta m^2_{ij}$ ($L\sim L_{0,ij}^\text{osc}$). The typical
values of $L/E$ for different types of neutrino sources and
experiments are summarized in Table \ref{tab:nuexp}.

\begin{table}[ht]
\caption{Characteristic values of $L$ and $E$ for various neutrino
  sources and experiments.}
\label{tab:nuexp}
\centering
\begin{tabular}{@{}>{\rule[-.2em]{0pt}{1.3em}}lccc@{}} \hline\hline
Experiment
   & $L~(\UmZ)$  & $E~(\UMeVZ)$ & $\Delta m^2~(\UeVZ^2)$    \\\hline
Solar
   & $10^{10}$   & $1$          & $10^{-10}$                \\
Atmospheric
   & $10^4$--$10^7$ & $10^2$--$10^5$  & $10^{-1}$--$10^{-4}$         \\
Reactor
   & $10^2$--$10^3$ & $1$             & $10^{-2}$--$10^{-3}$         \\
KamLAND
   & $10^5$         & $1$             & $10^{-5}$                    \\
Accelerator
   & $10^2$         & $10^3$--$10^4$  & $\gtrsim10^{-1}$             \\
Long-baseline accelerator
   & $10^5$--$10^6$ & $10^4$          & $10^{-2}$--$10^{-3}$         \\\hline\hline
\end{tabular}
\end{table}

If $(E/L)\gg\Delta m^2_{ij}$ ($L\ll L_{0,ij}^\text{osc}$), the
oscillation does not have time to give an appreciable effect because
$\sin^2x_{ij}\ll1$. The case of $(E/L)\ll\Delta m^2_{ij}$ ($L\gg
L_{0,ij}^\text{osc}$) requires more careful consideration. One  must
take into account that, in general, neutrino beams are not
monochromatic. Thus, rather than measuring $P_{\alpha\beta}$, the
experiments are sensitive to the average probability
\begin{equation}
\langle P_{\alpha\beta}\rangle=\delta_{\alpha\beta}
-4\sum_{i=1}^2\sum_{j=i+1}^3\mathcal{R}e \left(U_{\alpha i}U_{\beta
  i}^*U_{\alpha j}^*U_{\beta j}\right) \langle\sin^2 x_{ij}\rangle.
\end{equation}
For $L\gg L_{0,ij}^\text{osc}$, the oscillation phase goes through many
cycles before the detection and is averaged to $\langle\sin^2
x_{ij}\rangle=1/2$.

For a two-neutrino case,
\begin{equation}\label{nuvactwo}
P_{\alpha\beta}=\delta_{\alpha\beta}-
(2\delta_{\alpha\beta}-1)\sin^22\theta\sin^2x.
\end{equation}
For averaged oscillations we get, for example,
\begin{equation}
P_{ee}=1-\frac{1}{2}\sin^22\theta.
\end{equation}

For a recent careful derivation of the oscillation formulae, see
\Bref{Cohen:2008qb}.

\subsection{Neutrinos in matter}
\label{sec:mat}
When neutrinos propagate in dense matter, the interactions with the
medium affect their properties. These effects are either coherent or
incoherent. For purely incoherent $\nu$--$p$ scattering, the
characteristic cross-section is very small,
\begin{equation}\label{inccs}
\sigma\sim\frac{G_F^2s}{\pi}\sim10^{-43}\Ucm^2\left(\frac{E}{1\UMeV}\right)^2\SPp.
\end{equation}
The smallness of this cross-section is demonstrated by the fact that
if a beam of $10^{10}$ neutrinos with $E\sim1\UMeV$ was aimed at
Earth, only one would be deflected by the Earth's matter. It may seem
then that for neutrinos matter is irrelevant. However, one must take
into account that \Eref{inccs} does not contain the contribution
from forward elastic coherent interactions. In coherent interactions,
the medium remains unchanged and it is possible to have interference
of scattered and unscattered neutrino waves which enhances the
effect. Coherence further allows one to decouple the evolution
equation of neutrinos from the equations of the medium. In this
approximation, the effect of the medium is described by an effective
potential which depends on the density and composition of the matter
\cite{Wolfenstein:1977ue}.

Consider, for example, the effective potential for $\nu_e$
induced by its charged-current interactions with electrons in matter:
\begin{equation}\label{efpoee}
V_C=\langle \nu_e|\int d^3x
H_C^{(e)}|\nu_e\rangle=\sqrt{2}G_FN_e.
\end{equation}
For $\overline{\nu_e}$ the sign of $V$ is reversed. The potential can
also be expressed in terms of the matter density $\rho$:
\begin{equation}
V_C=7.6\ \frac{N_e}{N_p+N_n}\ \frac{\rho}{10^{14}\ \text{g/cm}^3}\UeV\SPp.
\end{equation}
Two examples that are relevant to observations are the following:
\begin{itemize}
  \item At the Earth's core $\rho\sim10\Ug/\UcmZ^3$ and
    $V\sim10^{-13}\UeV$.
\item At the solar core $\rho\sim100\Ug/\UcmZ^3$ and
  $V\sim10^{-12}\UeV$.
\end{itemize}

Consider a state that is an admixture of two neutrino species,
$|\nu_e\rangle$ and $|\nu_a\rangle$ or, equivalently,  $|\nu_1\rangle$
and $|\nu_2\rangle$. With some approximations, the time evolution
can be written in the following matrix form \cite{Wolfenstein:1977ue}:
\begin{equation}
 -i\frac{\partial}{\partial x} \begin{pmatrix}\nu_e\\\nu_a\end{pmatrix}
 =-\frac{1}{2E}M_w^2           \begin{pmatrix}\nu_e\\\nu_a\end{pmatrix}\SPp,
\end{equation}
where we have defined an effective mass matrix in matter,
\begin{equation}\label{hweaknu}
M_w^2=\frac{1}{2}
  \begin{pmatrix}
    m_1^2+m_2^2+4EV_e-\Delta m^2\cos2\theta &\Delta m^2\sin2\theta\\
   \Delta m^2\sin2\theta& m_1^2+m_2^2+4EV_a+\Delta m^2\cos2\theta
  \end{pmatrix}\SPp,
\end{equation}
with $\Delta m^2=m_2^2-m_1^2$.

We define the instantaneous mass eigenstates in matter, $\nu_i^m$, as
the eigenstates of $M_w$ for a fixed value of $x$. They are related to
the interaction eigenstates by a unitary transformation,
\begin{equation}
\begin{pmatrix}\nu_e                    \\ \nu_a\end{pmatrix}=U(\theta_m)
\begin{pmatrix}\nu_1^m                  \\ \nu_2^m\end{pmatrix}=
\begin{pmatrix}\cos\theta_m&\sin\theta_m\\ -\sin\theta_m&\cos\theta_m\end{pmatrix}
\begin{pmatrix}\nu_1^m                  \\ \nu_2^m\end{pmatrix}\SPp.
\end{equation}
The eigenvalues of $M_w$, that is, the effective masses in matter, are
given by \cite{Wolfenstein:1977ue,mism}
\begin{equation}
\mu^2_{1,2}=\frac{m_1^2+m_2^2}{2}+E(V_e+V_a)\mp\frac{1}{2}\sqrt{
  (\Delta m^2\cos2\theta-A)^2+(\Delta m^2\sin2\theta)^2},
\end{equation}
while the mixing angle in matter is given by
\begin{equation}
\tan2\theta_m=\frac{\Delta m^2\sin2\theta}{\Delta m^2\cos2\theta-A},
\end{equation}
where
\begin{equation}\label{defa}
A\equiv2E(V_e-V_a).
\end{equation}

The instantaneous mass eigenstates $\nu_i^m$ are, in general, not
energy eigenstates: they mix in the evolution. The importance of this
effect is controlled by the relative size of $4E\dot\theta_m(t)$ with
respect to $\mu_2^2(t)-\mu_1^2(t)$. When the latter is much larger
than the first, $\nu_i^m$ behave approximately as energy eigenstates
and do not mix during the evolution. This is the adiabatic transition
approximation. The adiabaticity condition reads
\begin{equation}\label{adicon}
\mu_2^2(t)-\mu_1^2(t)\gg 2EA\Delta m^2\sin2\theta\left|\dot
  A/A\right|.
\end{equation}
The transition probability for the adiabatic case is given by
\begin{equation}\label{peeadi}
P_{ee}(t)=\left|\sum_i U_{ei}(\theta)U_{ei}^*(\theta_p)\exp\left(-
    \frac{i}{2E}\int_{t_0}^t\mu_i^2(t^\prime)dt^\prime\right)\right|^2,
\end{equation}
where $\theta_p$ is the mixing angle at the production point. For
the case of two-neutrino mixing, \Eref{peeadi} takes the form
\begin{equation}\label{peeadtwo}
P_{ee}(t)=\cos^2\theta_p\cos^2\theta+\sin^2\theta_p\sin^2\theta
+\frac{1}{2}\sin2\theta_p\sin2\theta\cos\left(\frac{\delta(t)}{2E}\right),
\end{equation}
where
\begin{equation}
\delta(t)=\int_{t_p}^t[\mu_2^2(t^\prime)-\mu_1^2(t^\prime)]dt^\prime.
\end{equation}
For $\mu_2^2(t)-\mu_1^2(t)\gg E$, the last term in
\Eref{peeadtwo} is averaged out and the survival probability
takes the form
\begin{equation}\label{peeadifin}
P_{ee}=\frac{1}{2}[1+\cos2\theta_p\cos2\theta].
\end{equation}

The relative importance of the MSW matter term [$A$ of \Eref{defa}]
and the kinematic vacuum oscillation term in the Hamiltonian [the
off-diagonal term in \Eref{hweaknu}] can be parametrized by the
quantity $\beta_\text{MSW}$, which represents the ratio of matter to
vacuum effects (see, for example, \Bref{Bahcall:2004mz}).  From
\Eref{hweaknu} we see that the appropriate ratio is
\begin{equation}\label{defbeta}
\beta_\text{MSW}=\frac{2\sqrt{2}G_F n_e E_\nu}{\Delta m^2}.
\end{equation}
The quantity $\beta_\text{MSW}$ is the ratio between the oscillation length in
matter and the oscillation length in vacuum. In convenient units,
$\beta_\text{MSW}$ can be written as
\begin{equation}\label{bequan}
\beta_\text{MSW}=
0.19\left(\frac{E_\nu}{1\UMeV}\right)
    \left(\frac{\mu_e\rho}{100\Ug\Ucm^{-3}}\right)
    \left(\frac{8\times10^{-5}\UeV^2}{\Delta m^2}\right)\SPp.
\end{equation}
Here $\mu_e$ is the electron mean molecular weight
($\mu_e\approx0.5(1+X)$, where $X$ is the mass fraction of hydrogen)
and $\rho$ is the total density. If $\beta_\text{MSW}\lesssim\cos2\theta$,
the survival probability corresponds to vacuum averaged oscillations
[see \Eref{nuvactwo}],
\begin{equation}
P_{ee}=\left(1-\frac{1}{2} \sin^22\theta\right)\ \ \ (\beta_\text{MSW}<\cos2\theta,\
\text{vacuum}).
\end{equation}
If $\beta_\text{MSW}>1$, the survival probability corresponds to
matter-dominated oscillations [see \Eref{peeadifin}],
\begin{equation}
P_{ee}=\sin^2\theta\ \ \ (\beta_\text{MSW}>1,\ \text{MSW}).
\end{equation}
The survival probability is approximately constant in either of the
two limiting regimes, $\beta_\text{MSW}<\cos2\theta$ and
$\beta_\text{MSW}>1$. There is a strong energy dependence only in the
transition region between the limiting regimes.

For the Sun, $N_e(R)=N_e(0)\exp(-R/r_0)$, with $r_0\equiv
R_\odot/10.54=6.6\times10^7\ \text{m}=3.3\times10^{14}\UeV^{-1}$.
Then, the adiabaticity condition for the Sun reads
\begin{equation}
\frac{(\Delta
  m^2/\UeVZ^2)\sin^22\theta}{(E/\UMeVZ)\cos2\theta}\gg3\times10^{-9}.
\end{equation}



\begin{thebibliography}{99}

\bibitem{Kobayashi:1973fv} M.~Kobayashi and T.~Maskawa,
Prog.\ Theor.\ Phys.\  \textbf{49}, 652 (1973).


\bibitem{Cabibbo:1963yz} N.~Cabibbo,
Phys.\ Rev.\ Lett.\  \textbf{10}, 531 (1963).

\bibitem{Carter:1980hr}
  A.~B.~Carter and A.~I.~Sanda,
  Phys.\ Rev.\ Lett.\  \textbf{45}, 952 (1980);
  Phys.\ Rev.\  D \textbf{23}, 1567 (1981).

\bibitem{Bigi:1981qs}
  I.~I.~Y.~Bigi and A.~I.~Sanda,
  Nucl.\ Phys.\  B \textbf{193}, 85 (1981).

\bibitem{Buchalla:1995vs}
G.~Buchalla, A.~J.~Buras, and M.~E.~Lautenbacher,
Rev.\ Mod.\ Phys.\  \textbf{68}, 1125 (1996)
[arXiv:hep-ph/9512380].

\bibitem{Grossman:2002bu}
Y.~Grossman, A.~L.~Kagan, and Z.~Ligeti,
Phys.\ Lett.\ B \textbf{538}, 327 (2002)
[arXiv:hep-ph/0204212].

\bibitem{Boos:2004xp}
  H.~Boos, T.~Mannel, and J.~Reuter,
  Phys.\ Rev.\ D \textbf{70}, 036006 (2004)
  [arXiv:hep-ph/0403085].

\bibitem{Li:2006vq}
  H.~n.~Li and S.~Mishima,
  JHEP {\bf 0703}, 009 (2007)
  [arXiv:hep-ph/0610120].

\bibitem{Gronau:2008cc}
  M.~Gronau and J.~L.~Rosner,
  Phys.\ Lett.\  B {\bf 672}, 349 (2009)
  [arXiv:0812.4796 [hep-ph]].


\bibitem{hfag}
  E. Barberio \etal{} [Heavy Flavor Averaging Group],  
  arXiv:0808.1297 [hep-ex],
  online update at http://www.slac.stanford.edu/xorg/hfag

\bibitem{Yao:2006px}
  C.~Amsler \emph{et al.}  [Particle Data Group],
  Phys.\ Lett.\  B \textbf{667}, 1 (2008).

\bibitem{ckmfitter}
CKMfitter Group (J. Charles \etal),
Eur. Phys. J. C {\bf 41}, 1--131 (2005), [hep-ph/0406184],
updated results and plots available at: http://ckmfitter.in2p3.fr

\bibitem{Nir:2002gu}
  Y.~Nir,
  Nucl.\ Phys.\ Proc.\ Suppl.\  \textbf{117}, 111 (2003)
  [arXiv:hep-ph/0208080].

\bibitem{Grossman:1997dd}
Y.~Grossman, Y.~Nir, and M.~P.~Worah,
Phys.\ Lett.\ B \textbf{407}, 307 (1997) [hep-ph/9704287].

\bibitem{Grossman:2006ce}
  Y.~Grossman, Y.~Nir, and G.~Raz,
  Phys.\ Rev.\ Lett.\  \textbf{97}, 151801 (2006)
  [arXiv:hep-ph/0605028].

\bibitem{Bona:2007vi}
  M.~Bona \emph{et al.}  [UTfit Collaboration],
  JHEP \textbf{0803}, 049 (2008)
  [arXiv:0707.0636 [hep-ph]].

\bibitem{Branco:1999fs}
G.~C.~Branco, L.~Lavoura, and J.~P.~Silva,
\emph{CP Violation}
(Clarendon Press, Oxford, 1999).

\bibitem{Bigi:2000wn}
  I.~I.~Y.~Bigi and N.~G.~Uraltsev,
  Nucl.\ Phys.\  B \textbf{592}, 92 (2001)
  [arXiv:hep-ph/0005089].

\bibitem{Falk:2001hx}
  A.~F.~Falk, Y.~Grossman, Z.~Ligeti, and A.~A.~Petrov,
  Phys.\ Rev.\  D \textbf{65}, 054034 (2002)
  [arXiv:hep-ph/0110317].

\bibitem{Falk:2004wg}
  A.~F.~Falk, Y.~Grossman, Z.~Ligeti, Y.~Nir, and A.~A.~Petrov,
  Phys.\ Rev.\  D \textbf{69}, 114021 (2004)
  [arXiv:hep-ph/0402204].

\bibitem{Aubert:2007wf}
  B.~Aubert \emph{et al.}  [BaBar Collaboration],
  Phys.\ Rev.\ Lett.\  \textbf{98}, 211802 (2007)
  [arXiv:hep-ex/0703020].

\bibitem{Staric:2007dt}
  M.~Staric \emph{et al.}  [Belle Collaboration],
  Phys.\ Rev.\ Lett.\  \textbf{98}, 211803 (2007)
  [arXiv:hep-ex/0703036].


\bibitem{Raz:2002zx}
  G.~Raz,
  Phys.\ Rev.\ D \textbf{66}, 037701 (2002)
  [arXiv:hep-ph/0205310].

\bibitem{ArkaniHamed:2004fb}
  N.~Arkani-Hamed and S.~Dimopoulos,
  JHEP \textbf{0506}, 073 (2005)
  [arXiv:hep-th/0405159].

\bibitem{Cohen:1996vb}
  A.~G.~Cohen, D.~B.~Kaplan, and A.~E.~Nelson,
  Phys.\ Lett.\  B \textbf{388}, 588 (1996)
  [arXiv:hep-ph/9607394].

\bibitem{Nir:2002ah}
  Y.~Nir and G.~Raz,
  Phys.\ Rev.\  D \textbf{66}, 035007 (2002)
  [arXiv:hep-ph/0206064].

\bibitem{Blum:2009sk}
  K.~Blum, Y.~Grossman, Y.~Nir and G.~Perez,
  Phys.\ Rev.\ Lett.\  {\bf 102}, 211802 (2009)
  [arXiv:0903.2118 [hep-ph]].

\bibitem{Nir:1993mx}
Y.~Nir and N.~Seiberg,
Phys.\ Lett.\ B \textbf{309}, 337 (1993)
[arXiv:hep-ph/9304307].

\bibitem{Leurer:1993gy}
M.~Leurer, Y.~Nir, and N.~Seiberg,
Nucl.\ Phys.\ B \textbf{420}, 468 (1994)
[arXiv:hep-ph/9310320].

\bibitem{Ciuchini:2007cw}
  M.~Ciuchini, E.~Franco, D.~Guadagnoli, V.~Lubicz, M.~Pierini, V.~Porretti, and L.~Silvestrini,
  Phys.\ Lett.\  B \textbf{655}, 162 (2007)
  [arXiv:hep-ph/0703204].

\bibitem{Nir:2007ac}
  Y.~Nir,
  JHEP \textbf{0705}, 102 (2007)
  [arXiv:hep-ph/0703235].

\bibitem{D'Ambrosio:2002ex}
  G.~D'Ambrosio, G.~F.~Giudice, G.~Isidori, and A.~Strumia,
  Nucl.\ Phys.\  B \textbf{645}, 155 (2002)
  [arXiv:hep-ph/0207036].

\bibitem{Grossman:2007bd}
  Y.~Grossman, Y.~Nir, J.~Thaler, T.~Volansky, and J.~Zupan,
  Phys.\ Rev.\  D \textbf{76}, 096006 (2007)
  [arXiv:0706.1845 [hep-ph]].

\bibitem{Feng:2007ke}
  J.~L.~Feng, C.~G.~Lester, Y.~Nir, and Y.~Shadmi,
  Phys.\ Rev.\  D \textbf{77}, 076002 (2008)
  [arXiv:0712.0674 [hep-ph]].

\bibitem{Engelhard:2009br}
  G.~Engelhard, J.~L.~Feng, I.~Galon, D.~Sanford and F.~Yu,
  arXiv:0904.1415 [hep-ph].

\bibitem{Feng:2009bs}
  J.~L.~Feng, I.~Galon, D.~Sanford, Y.~Shadmi and F.~Yu,
  Phys.\ Rev.\  D {\bf 79}, 116009 (2009)
  [arXiv:0904.1416 [hep-ph]].

\bibitem{Feng:2009yq}
  J.~L.~Feng, S.~T.~French, C.~G.~Lester, Y.~Nir and Y.~Shadmi,
  arXiv:0906.4215 [hep-ph].
  
\bibitem{Hiller:2008wp}
  G.~Hiller and Y.~Nir,
  JHEP \textbf{0803}, 046 (2008)
  [arXiv:0802.0916 [hep-ph]].

\bibitem{Hiller:2008sv}
  G.~Hiller, Y.~Hochberg, and Y.~Nir,
  arXiv:0812.0511 [hep-ph].

\bibitem{Nomura:2007ap}
  Y.~Nomura, M.~Papucci, and D.~Stolarski,
  Phys.\ Rev.\  D \textbf{77}, 075006 (2008)
  [arXiv:0712.2074 [hep-ph]];
  JHEP \textbf{0807}, 055 (2008)
  [arXiv:0802.2582 [hep-ph]].

\bibitem{HiHo2009}
  G.~Hiller, Y.~Hochberg, and Y.~Nir, work in progress.

\bibitem{Giudice:2008uk}
  G.~F.~Giudice, M.~Nardecchia, and A.~Romanino,
  arXiv:0812.3610 [hep-ph].

\bibitem{Nelson:2000sn}
  A.~E.~Nelson and M.~J.~Strassler,
  JHEP \textbf{0009}, 030 (2000)
  [arXiv:hep-ph/0006251];
  JHEP \textbf{0207}, 021 (2002)
  [arXiv:hep-ph/0104051].

\bibitem{Gonzalez-Garcia:2002dz}
  M.~C.~Gonzalez-Garcia and Y.~Nir,
  Rev.\ Mod.\ Phys.\  \textbf{75}, 345 (2003)
  [arXiv:hep-ph/0202058].

\bibitem{GonzalezGarcia:2007ib}
  M.~C.~Gonzalez-Garcia and M.~Maltoni,
  Phys.\ Rep.\  \textbf{460}, 1 (2008)
  [arXiv:0704.1800 [hep-ph]].

\bibitem{Fukugita:1986hr}
  M.~Fukugita and T.~Yanagida,
  Phys.\ Lett.\  B \textbf{174}, 45 (1986).

\bibitem{Davidson:2008bu}
  S.~Davidson, E.~Nardi, and Y.~Nir,
  Phys.\ Rep.\  \textbf{466}, 105 (2008)
  [arXiv:0802.2962 [hep-ph]].

\bibitem{Davidson:2002qv}
  S.~Davidson and A.~Ibarra,
  Phys.\ Lett.\  B \textbf{535}, 25 (2002)
  [arXiv:hep-ph/0202239].

\bibitem{Harrison:2002er}
  P.~F.~Harrison, D.~H.~Perkins, and W.~G.~Scott,
  Phys.\ Lett.\  B \textbf{530}, 167 (2002)
  [arXiv:hep-ph/0202074].

\bibitem{Froggatt:1978nt}
  C.~D.~Froggatt and H.~B.~Nielsen,
  Nucl.\ Phys.\  B \textbf{147}, 277 (1979).

\bibitem{Leurer:1992wg}
  M.~Leurer, Y.~Nir, and N.~Seiberg,
  Nucl.\ Phys.\  B \textbf{398}, 319 (1993)
  [arXiv:hep-ph/9212278].

\bibitem{Grossman:1995hk}
  Y.~Grossman and Y.~Nir,
  Nucl.\ Phys.\  B \textbf{448}, 30 (1995)
  [arXiv:hep-ph/9502418].

\bibitem{Hall:1999sn}
  L.~J.~Hall, H.~Murayama, and N.~Weiner,
  Phys.\ Rev.\ Lett.\  \textbf{84}, 2572 (2000)
  [arXiv:hep-ph/9911341].

\bibitem{Antebi:2005hr}
  Y.~E.~Antebi, Y.~Nir, and T.~Volansky,
  Phys.\ Rev.\  D \textbf{73}, 075009 (2006)
  [arXiv:hep-ph/0512211].

\bibitem{Chau:1984fp}
L.~Chau and W.~Keung,
Phys.\ Rev.\ Lett.\  \textbf{53}, 1802 (1984).

\bibitem{Wolfenstein:1983yz}
L.~Wolfenstein,
Phys.\ Rev.\ Lett.\  \textbf{51}, 1945 (1983).

\bibitem{Buras:1994ec}
A.~J.~Buras, M.~E.~Lautenbacher, and G.~Ostermaier,
Phys.\ Rev.\ D \textbf{50}, 3433 (1994)
[arXiv:hep-ph/9403384].

\bibitem{Dib:1989uz}
  C.~Dib, I.~Dunietz, F.~J.~Gilman, and Y.~Nir,
  Phys.\ Rev.\ D \textbf{41}, 1522 (1990).

\bibitem{Rosner:1988nx}
  J.~L.~Rosner, A.~I.~Sanda, and M.~P.~Schmidt,
EFI-88-12-CHICAGO
[Presented at Workshop on High Sensitivity Beauty Physics, Batavia, IL, Nov 11--14, 1987].

\bibitem{Weisskopf:au}
V.~Weisskopf and E.~P.~Wigner,
Z.\ Phys.\  \textbf{63}, 54 (1930);
Z.\ Phys.\  \textbf{65}, 18 (1930).
[See Appendix A of P.~K.~Kabir, \emph{The CP Puzzle: Strange Decays of the
Neutral Kaon} (Academic Press, London, 1968).]

\bibitem{Nir:1992uv}
  Y.~Nir,
SLAC-PUB-5874
[Lectures given at \emph{20th Summer Institute on Particle Physics:
The Third Family and the Physics of Flavor}, Stanford, CA, 1992,
ed. L. Vassilian (SLAC, Stanford, 1993)].

\bibitem{Dunietz:1986vi}
I.~Dunietz and J.~L.~Rosner,
Phys.\ Rev.\ D \textbf{34}, 1404 (1986).

\bibitem{Blinov:ru}
Ya.~I.~Azimov, N.~G.~Uraltsev, and V.~A.~Khoze,
Sov.\ J.\ Nucl.\ Phys.\  \textbf{45}, 878 (1987)
[Yad.\ Fiz.\  \textbf{45}, 1412 (1987)].

\bibitem{Bigi:1986vr}
I.~I.~Bigi and A.~I.~Sanda,
Nucl.\ Phys.\ B \textbf{281}, 41 (1987).

\bibitem{Pontecorvo:1957cp}
  B.~Pontecorvo,
  Sov.\ Phys.\ JETP \textbf{6}, 429 (1957)
  [Zh.\ Eksp.\ Teor.\ Fiz.\  \textbf{33}, 549 (1957)].

\bibitem{Cohen:2008qb}
  A.~G.~Cohen, S.~L.~Glashow, and Z.~Ligeti,
  arXiv:0810.4602 [hep-ph].

\bibitem{Wolfenstein:1977ue}
  L.~Wolfenstein,
  Phys.\ Rev.\  D \textbf{17}, 2369 (1978).

\bibitem{mism}
  S.P. Mikheyev and A. Yu. Smirnov,
  Sov.\ J. Nucl. Phys.\ \textbf{42}, 913 (1985)
  [Yad.\ Fiz.\  \textbf{42}, 1441 (1985)].

\bibitem{Bahcall:2004mz}
  J.~N.~Bahcall and C.~Pena-Garay,
  New J.\ Phys.\  \textbf{6}, 63 (2004)
  [arXiv:hep-ph/0404061].


\end{thebibliography}
\end{document}